\documentclass[]{aa} % for the long lists of affiliations 
\usepackage{multirow}
\usepackage{caption}
\usepackage{adjustbox}
\usepackage{subcaption}
\usepackage{graphicx}
\usepackage[varg]{txfonts}
\usepackage{natbib}
\usepackage{hyperref}
\usepackage[utf8]{inputenc}
\usepackage{xcolor}
\usepackage{lscape}
\usepackage{stfloats}
\usepackage{cleveref}
\usepackage[table,xcdraw]{xcolor}
\usepackage[T1]{fontenc}     % 
\usepackage{placeins}
\usepackage{babel}           % For language-specific typographic rules
\usepackage{graphicx}
\usepackage{float} % Permite usar [H]
\usepackage[normalem]{ulem}

\usepackage{comment}

\newcommand{\vnh}{\hat{\mathbf n}}

\usepackage{hyperref}
\hypersetup{
  colorlinks,
  filecolor=blue,
  citecolor=blue,
  linkcolor=blue,
  urlcolor=blue}

\begin{document} 
\title{Towards precision cosmology with Voids $\times$ CMB correlations (I)}
\subtitle{{\emph{Roman}-\textsc{Agora}} mock catalogs and pipeline validation}

\titlerunning{Towards precision cosmology with Roman Void $\times$ CMB correlations}
\author{%
Mar Pérez Sar\inst{1,2}\thanks{\href{mailto:perezsarmar@gmail.com}{perezsarmar@gmail.com} }
\and Carlos Hernández Monteagudo\inst{1}
\and András Kovács\inst{3,4}
\and Alice Pisani\inst{5,6}}

\institute{%
Instituto de Astrofísica de Canarias, Calle Vía Láctea s/n, E-38205, La Laguna, Tenerife, Spain
\and
Departamento de Astrofísica, Universidad de La Laguna, E-38206, La Laguna, Tenerife, Spain
\and
MTA–CSFK Lendület “Momentum” Large-Scale Structure (LSS) Research Group, Konkoly Thege Miklós út 15-17, H-1121 Budapest, Hungary
\and
Konkoly Observatory, HUN-REN Research Centre for Astronomy and Earth Sciences, Budapest, Hungary
\and
CPPM, Aix-Marseille Université, CNRS/IN2P3, Marseille, France
\and
Department of Astrophysical Sciences, Peyton Hall, Princeton University, Princeton, NJ 08544, USA
}

   \date{Received xxx, 2026; accepted xxx, 2026}

% \abstract{}{}{}{}{} 
% 5 {} token are mandatory
%  \abstract


\abstract{The Nancy Grace \emph{Roman} Space Telescope, in combination with forthcoming CMB experiments, will enable high-precision cross-correlations such as Void $\times$ CMB lensing. However, extracting cosmological information from these measurements requires modeling the signal’s dependence on cosmology and characterizing methodological and mock-dependent systematics. This motivates the need of realistic mock catalogs with integrated CMB maps.
To this end, we construct and validate a suite of multi-purpose mock galaxy catalogs with attached CMB maps designed to approximate, with varying levels of fidelity, the main characteristics of the \emph{Roman} survey. Our approach differs from traditional halo-occupation or abundance-matching methods by directly translating a reference mock catalog—containing basic host-halo properties—into a new simulation (here \textsc{Agora}). This technique, which we call \emph{analog matching}, assigns each reference galaxy a halo counterpart in the new simulation via a nearest-neighbor search in a multi-dimensional parameter space (including  halo mass, environmental measures, and other galaxy attributes). By varying this parameter vector, we generate catalogs of differing complexity and systematically test how different galaxy-halo prescriptions influence large-scale structure (LSS) statistics and CMB-related observables.
We find that \emph{analog matching} based on halo mass alone, or halo mass with galaxy-type indicators, successfully reproduces the expected \emph{Roman} emission-line galaxy statistics. However, reproducing two-dimensional galaxy clustering does not guarantee consistent void properties. Our results highlight that void measurements provide an independent and sensitive test of galaxy-halo modeling beyond the matter power spectrum.
An important by-product of our setup is that it is fully general and can be applied to any combination of simulation and reference catalog, given that the parameter space for both is specified. The resulting \emph{Roman}-\textsc{Agora} mock catalogs offer a versatile tool for LSS $\times$ CMB studies and for testing how modeling impacts cosmological observables.}


   \maketitle \nolinenumbers

\section{Introduction}
\label{sec:intro}

From our tiny blue point of view, we have expanded our understanding of the Universe by collecting vast datasets across the electromagnetic spectrum and cosmic history. Interpreting these data through theoretical models and simulations has allowed us to test and establish the $\Lambda$ cold dark matter model ($\Lambda$CDM) as our standard cosmological model. This model only requires six parameters, general relativity and linear perturbations around a homogeneous and isotropic background to successfully describe most observations in a wide range of scales. Yet, its main ingredients, dark matter and dark energy, remain unknown, and as measurements and analysis techniques improve, subtle but persistent tensions have emerged. This suggests the presence of unaccounted systematic effects or new physics beyond our current understanding. \citep{LCDMtensionsreview_2022jun_abdalla,LCDMtensionsreview_2022dec_perivolaropoulos,LCDMtensionsreview_2022dec_peebles,LCDMtensionsreview_2025feb_efstathiou}. 

Cross-correlating independent cosmological probes has become an essential strategy to address these tensions. Joint analyses exploit the complementary sensitivities of different observables, break degeneracies between cosmological parameters, mitigate systematics errors unique to each individual dataset, and enhance the signal-to-noise of measurements that would otherwise remain undetectable \citep{potentialofCMBxLSS_2008aug_ho,potentialofCMBxLSS_2008aug_hirata,potentialofCMBxLSS_2014feb_pearson,potentialofCMBxLSS_2016oct_nicola,potentialofCMBxLSS_DESjan23_omori,potentialofCMBxLSS_DESjan23_chang,potentialofCMBxLSS_DESjan23_abbot}.

The cross-correlation between the Cosmic Microwave Background (CMB) and the Large-Scale Structure (LSS) is particularly powerful. As CMB photons travel through the evolving matter distribution, their trajectories and energies are perturbed by both gravitational and scattering processes. The first arise from the large-scale gravitational potential, where spatial variations induce gravitational lensing \citep{lewisandchallinor2006} and temporal evolution gives rise to the Integrated Sachs–Wolfe (ISW) effect \citep{sachs&wolfe1967}. In parallel, scattering processes happen as CMB photons interact with the baryonic constituents of the LSS, including scattering off free electrons via the kinetic and thermal Sunyaev–Zel’dovich effects (kSZ and tSZ, respectively) \citep{kSZ_sunyaev_80,tSZ_sunyaev_72}, neutral hydrogen via Rayleigh scattering \citep{basu_2004,lewis_rayleigh_13}, and metallic or ionic species through (sub-)millimeter fine-structure transitions \citep{basu_2004,chm_metals_reio,chm_metals_pol}. All these LSS-induced distortions reprocess the original CMB signal, imprinting late-time information about the growth of structure and the geometry of the Universe. The resulting CMB $\times$ LSS signals provide constraints on cosmological parameters such as the matter density ($\Omega_{\rm m}$), clustering amplitude ($\sigma_8$), neutrino masses, and dark energy, \citep{cosmological_constraints_fromCMBxLSS_2021may_Chen,cosmological_constraints_fromCMBxLSS_2021dec_krolewski,cosmological_constraints_fromCMBxLSS_2024feb_shaikh} and reveal important systematic effects that can bias the cosmological inference if not accounted for. Examples include the contamination of CMB lensing reconstructions by thermal SZ foregrounds and other higher-order LSS correlations \citep{contaminationCMBlensing_2014mar_troxel}.

Realistic simulations are essential for extracting cosmological information from CMB $\times$ LSS measurements. Although analytical and semi-analytical models exist, they rely on simplifying assumptions that often break down in the non-linear regime and under realistic observational conditions. Simulations, therefore, provide the only framework capable of capturing the non-linear evolution of structure and the correlated physics that produce the observed CMB $\times$ LSS signals. They allow us to model structure formation self-consistently, test and validate analysis pipelines, estimate uncertainties and covariances, and identify potential systematic biases. 

Building simulations that coherently include CMB and LSS observables is, however, technically demanding. LSS tracers require high mass and spatial resolution to resolve small-scale structures, while CMB observables demand very large volumes to capture long-wavelength modes and line-of-sight projections. Moreover, interdependencies between different signals---namely CMB lensing, the SZ effects, and the Cosmic Infrared Background (CIB)---must be modeled consistently to prevent biased cross-correlations. Meeting these requirements imposes heavy computational and storage costs and limits the number of independent realizations that can be produced.

Some examples of simulations that try to overcome these challenges are \textsc{WebSky} \citep{Websky2020}, \textsc{DEMNUni} \citep{DEMNUni2015Jcastorina,DEMNUni2016carbone}, \textsc{Agora} \citep{omori2022}, \textsc{Gower Street} \citep{gowerstreet2025}, and \textsc{HalfDome} \citep{halfdome2025}, each adopting different compromises between resolution, number of realizations and cosmological coverage. Despite their specific trade-offs, all these simulations are highly valuable for cross-correlation studies and the optimal choice depends on the scientific application.

The last step, once a simulation is selected, is to make it suitable for cosmological analysis by tailoring its data products to the specifications of the target survey. CMB-related outputs---including lensing, ISW, kSZ, tSZ, and CIB maps---typically require minimal processing. These maps are adjusted by applying the survey mask or beam, adding noise realizations consistent with the instrumental sensitivity, and, if necessary, applying small multiplicative calibrations (typically at the few-percent level) to match the observed power spectra. Their validation is performed statistically, through comparisons of auto- and cross-power spectra with theoretical predictions (for example, from \texttt{CAMB}\footnote{\hyperlink{https://camb.readthedocs.io/}{https://camb.readthedocs.io/}}) and with existing measurements from experiments such as \emph{Planck} \citep{Planck2014_survey,Planck2016_survey,Planck2020_survey}, the Atacama Cosmology Telescope (ACT) \citep{ACT2011_survey,ACT2017_survey} , or the South Pole Telescope (SPT) \citep{SPT2015_survey,SPT2019_survey}.

In contrast, the LSS products---which at this stage describe only the dark matter distribution and its associated halos---require more extensive adaptation to emulate the galaxy surveys. Halos must be populated with galaxies using empirical or semi-analytic prescriptions that account for the survey’s geometry, selection function, and galaxy bias \citep{somerville2014,wechsler&tinker2018}. This bias captures the complex mapping shaped by baryonic processes, local environment, and halo assembly history that modulate the resulting galaxy clustering relative to the underlying dark matter. 

Mock catalogs are usually validated from the galaxy clustering perspective---ensuring that they reproduce the one- and two-point statistics of the target survey---but they are rarely tested with higher-order statistics. Cosmic voids provide such a complementary test.

Cosmic voids form the low-density counterparts to the bright network that shape the cosmic web \citep{joever&einasto1978,gregory&thompson1978,pisani2019review, moresco2022review}. They arise in regions where matter is gravitationally drained towards surrounding overdensities, leading to interiors that reach only a fraction of the mean cosmic density. Importantly, voids extend across tens to hundreds of megaparsecs and occupy most of the Universe’s volume, evolving in a regime where dynamics remain close to linear and baryonic effects are minimal \citep{lavaux&wandelt2012_stacking_vgcf_aprsd,hamaus2014_dens_and_vel,paillas2017}. Their statistics encode higher-order correlations beyond traditional two-point clustering, providing additional means to evaluate how accurately mocks reproduce the full topology of the matter distribution \citep[see e.g.,][]{Arsenov2025}.

These properties make voids theoretically clean and highly sensitive cosmological probes, capable of constraining the expansion history, dark energy, and gravity \citep{
biswas2010_probe_de,bos2012_probe_de,pisani2015_vsf,
pollina2016_probe_de,hawken2017_vgcf_rsd1,
contarini2019_vsf,architouv2019_vgcf_rsd,correa2019_new_probe,nadathur2020_vgcf_aprsd,
paillas2021_vgcf_aprsd,woodfinden2022_vgcf_aprsd,pelliciari2023_vsf,
contarini2023_vsf,mauland2023_vgcf_aprsd,song2024_vsf,fraser2025_vgcf_aprsd}  primarily through the void size function  \citep{sheth&vdweygaert2004_vsf, furlanetto2006_vsf, jennings2013_vsf, pisani2015_vsf, ronconi2017_vsf, correa2021_vsf, contarini2023_vsf, verza2024_vsf, elena2025_vsf} and the void-galaxy cross-correlation function (with its distortions in redshift space) \citep{kaiser1987, alcock_1979, lavaux&wandelt2012_stacking_vgcf_aprsd, sutter2012_vgcf_ap, pisani2014_vgcf, cai2016_vgcf_rsd, mao2017_vgcf_ap, nadathur2019_vgcf_aprsd, aubert2022_vgcf_rsd, correa2022_vgcf_aprsd, hamaus_2022_vgcf_aprsd, woodfinden2022_vgcf_aprsd, radinovic2023_vgcf_aprsd, verza2024_vgcf_and_vsf_aprsd, degni2025_vgcf_aprsd,schuster2025_dens}.

In this work, we construct and validate a set of multi-purpose mock galaxy catalogs designed to capture, to different degrees of realism, the main characteristics of the Nancy Grace \emph{Roman} Space Telescope survey \footnote{\url{https://roman.gsfc.nasa.gov/}} \citep{ROMAN2015_survey}, as a foundation for void statistic and various CMB cross-correlation analyses. \emph{Roman}'s redshift coverage (1 $\leq z \leq$ 3) overlaps with other modern cosmology missions such as \emph{Euclid} \citep{EUCLID2011_survey} and the Dark Energy Spectroscopic Instrument (DESI) \citep{DESI2016_survey}, but its depth and area will provide a finer angular resolution and a higher sampling density for accessing smaller scales. This offers an interesting point of view for cosmic void studies, which so far have been investigated in wide-areas and sparser galaxy datasets.

For this purpose, we choose the \textsc{Agora} simulation \citep{omori2022} since its mass and spatial resolution is sufficient for resolving the low-mass halos that host emission-line galaxies (ELGs), while also provides a coherently suite of secondary CMB anisotropy maps and astrophysical foregrounds.

This \emph{Roman}-\textsc{Agora} mock framework offers a versatile tool for the community, enabling not only void analyses but also a broad range of future LSS-CMB cross-correlation studies, including 5×2pt, tSZ, and kSZ measurements. Although this paper focuses on describing and validating the mock catalogs themselves, a companion work will use these data products to investigate how different methodological choices---such as the mock catalog, void definitions, stacking techniques, and CMB map treatments---impact the Void $\times$ CMB lensing signal and its forecasted detectability with \emph{Roman} and upcoming CMB surveys.

The paper is organized as follows. In Section \ref{sec:datasets} we describe the data employed in this study. In Section \ref{sec:methodology}, we present the methodology and discuss its limitations. Section \ref{sec:results} contains the results and Section \ref{sec:conclusions} concludes with a summary of our main findings.

\section{Datasets}
\label{sec:datasets}
\subsection{\emph{Roman} Space Telescope}
The Nancy Grace \emph{Roman} Space Telescope (hereafter \emph{Roman}) is a NASA mission schedule for launch in late 2026, designed to conduct pioneering research on dark energy, exoplanets and infrared astronomy \citep{green2012wide,spergel2015wide,zasowski2025roman}. It has the same 2.4 m mirror as Hubble, but a field of view 100 times larger, enabling much faster wide-area surveys (1000 times faster than Hubble). 

Our analysis is based on the High Latitude Wide Area Survey (HLWAS), which combines imaging and spectroscopy to produce a 3-dimensional map of galaxies. The telescope's spectroscopy and imaging will operate in the near-infrared and visible wavelengths (only imaging) measuring millions of Emission-Line Galaxies (ELGs) as tracers of the underlying matter distribution. ELGs, identified through strong nebular emission lines such as \ion{H}{$\alpha$}, [\ion{O}{II}], and [\ion{O}{III}], arise primarily from active star formation and, to a lesser extent, AGN activity, allowing precise spectroscopic redshift measurements \citep{gonzalez2020}. 

The expected outcome of \emph{Roman}, predicted from reference catalogs in the literature is a total of 14.2 million \ion{H}{$\alpha$} redshifts, 3.6 million [\ion{O}{III}] redshifts, and 1.3 million [\ion{O}{II}] redshifts, with 11 million objects at $\rm z>1$ and 1.7 million at $\rm z>2$. Accurate redshift determination ideally requires the detection of two emission lines. When this is not possible, complementary imaging is necessary to help confirm the emission lines and to separate overlapping spectra. The grism’s wavelength range (1–1.9 $\mu$m) ensures reliable detection of \ion{H}{$\alpha$} (0.656 $\mu$m) and [\ion{O}{III}] (0.501 $\mu$m) for redshifts between $1.1 < z < 1.9$, and [\ion{O}{III}] and [\ion{O}{II}] (0.373 $\mu$m) for $1.8 < z < 2.8$. 

\emph{Roman} will carry out a three-tier survey: medium (2415 deg$^2$), wide (2700 deg$^2$), and deep (19.2 deg$^2$), covering 5100 deg$^2$ in 520 days \citep{zasowski2025roman}. Spectroscopy is limited to the medium and deep tiers ($\sim$2415 deg$^2$).

\subsubsection{\emph{Roman} mock catalogs}
To optimize the survey strategies and forecast its scientific outcome, several mock catalogs have been presented in the literature simulating \emph{Roman}'s performance---see e.g. \citet{wang2022} for predictions for baryon acoustic oscillations and redshift space distortions or \citet{verza2024cosmology} for cosmic void statistics---though none incorporate self-consistent CMB maps.

The existing \emph{Roman} mock catalog that we aim to replicate is presented in \cite{zhai2021clustering}\footnote{\url{https://irsa.ipac.caltech.edu/data/theory/Roman/Zhai2021/}}. It was created using a semi-analytical galaxy formation model (SAM) \textsc{Galacticus} \citep{benson2012galacticus}, coupled with the N-body simulation UNIT \citep[][see Table~\ref{tab:comparison_sims} for specifications]{chuang2019unit}. The SAM simulates the galaxy properties (such as stellar mass, star formation rate, metallicity, line luminosity) of the ELGs detected by the instrument in a $\sim$2000 deg$^2$ area, and calibrate them to match the observed \ion{H}{$\alpha$} luminosity function from the High-z Emission Line Survey (HiZELS, \citealt{geach2008hizels,sobral2009hizels,sobral2013large}). 

\subsection{\textsc{Agora} simulation}
To generate a \emph{Roman}-like mock catalog with associated CMB maps, we use the \textsc{Agora} simulation by \cite{omori2022}. It provides a single halo lightcone catalog with both CMB and LSS observables, and it includes CMB lensing convergence ($\kappa$), thermal and kinetic Sunyaev-Zel'dovich effects (tSZ/kSZ), the cosmic infrared background (CIB), radio sources, as well as galaxy density contrast and weak lensing maps.

\textsc{Agora} is based on the dark matter-only N-body simulation \textsc{MultiDark Planck 2} (MDPL2) \citep{klypin2016}\footnote{Description of the simulation and various data products can be accessed at \url{https://www.cosmosim.org/metadata/mdpl2/}, with halo catalogs available at \url{http://halos.as.arizona.edu/simulations/MDPL2/hlists/}.}, chosen for its extensive set of derived products that can be directly integrated with the simulation. Some examples are galaxy catalogs from semi-analytical models \citep{cora2018}, studies of emission line galaxy populations \citep{alam2021}, and intensity mapping predictions \citep{sato2023}. 

Appendix \ref{appendix:agora_details} provides a detailed description of how the LSS and CMB components of the \textsc{Agora} simulation are constructed. Appendix \ref{appendix:unit_vs_mdpl2} includes a comparison between \textsc{MDPL2} and the UNIT simulation (the parent simulations of the \emph{Roman} reference catalog and \textsc{Agora}, respectively) to justify our use of \textsc{Agora} for generating \emph{Roman}-like catalogs.

\section{Methodology}
\label{sec:methodology}

We review existing mock-generation methods to motivate and introduce our approach for the \emph{Roman} galaxy catalog. We then summarize different void definitions, justify the ones adopted in this work, and detail the pipelines used to construct the void catalogs.

\subsection{State of the art in mock catalogs}
As outlined in the introduction, mock galaxy catalogs are synthetic datasets simulating the universe's galaxy distribution, that play and essential role in modern large-scale structure analyses. Constructing these catalogs requires modeling the galaxy–halo connection, which links galaxies to the underlying dark matter distribution \citep{kaiser1984spatial,bardeen1986statistics,mo1996analytic,fry1996evolution,tegmark1998time}. Halo abundance and clustering are primarily determined by cosmology and are well captured in (pure dark matter) N-body simulations. Nonetheless, mapping from halos to galaxies is significantly more complex, since this process is heavily influenced by baryonic physics that cannot be modeled trivially, such as gas cooling, star formation, and AGN and supernova feedback.

Galaxy–halo connections can be modeled with physics-based or empirical-based approaches, tuned to reproduce fundamental observational quantities. Physics-based methods capture galaxy formation from first principles and include \textbf{hydrodynamical simulations}, which model the co-evolution of dark matter and baryons simultaneously, solving the fundamental equations of gravity, hydrodynamics, and thermodynamics  \citep{dubois2014,schaller2015,dolag2016,mccarthy2016bahamas,springel2018illustrisTNG,villaescusa2021camels,schaye2023flamingo}, 
and \textbf{semi-analytic models (SAMs)}, which use the information of halo merger trees from N-body simulations and simplified analytic recipes to model the relevant baryonic processes \citep{somerville2014,gabrielpillai2022,luciaperez2023,bower2006,behroozi2019universemachine,benson2012galacticus}. On the other hand, empirical methods derive statistical galaxy-halo mappings to match observations. They include \textbf{Sub-Halo Abundance Matching (SHAM)} methods that establish a monotonic relationship between a galaxy property, such as stellar mass or luminosity, and a property of its host dark matter halo, like its mass or peak velocity (the fundamental premise is that the most massive galaxy resides in the most massive halo) \citep{conroy2006,moster2010,behroozi2010,tacchella2013,moster2013,moster2018,tacchella2018}; and \textbf{Halo Occupation Distribution (HOD)} methods, which 
specify the probability distribution for a halo of a given mass to host a certain number of galaxies \citep{peacock2000,berlind2002,zheng2005,guo2015,rodriguez2015}. While the physical-driven methods are more precise, they are computationally expensive and not the primary choice for LSS studies. 

From the empirical-based category, even if simplistic, they have proven remarkably successful in reproducing a wide range of galaxy statistics, including clustering and lensing \citep{hearin2015dark}, 
with several key contributions over the last years. A major lesson from these studies, however, is that, while halo mass can be the leading property that links galaxies with halos \citep{white1978}, it cannot always by itself explain the observed clustering, especially on small scales \citep{yuan2021,yuan2022,Wu2024,Hadzhiyska2021,Paviot2024}. 

Halo occupation by galaxies can also depend on secondary halo properties, including assembly history---with earlier forming halos more clustered \citep{sheth2004,gao2005,wechsler2006,gao2007,dalal2008,angulo2008,li2008,faltenbacher2009,lazeyras2017,salcedo2018,han2019,sato2019,ramakrishnan2019,tucci2021,montero2021,balaguera2024}---, structural properties such as concentration and spin \citep{gao2007,faltenbacher2009,sato2019,montero2021}, and environmental factors \citep{kauffmann2004,hahn2006,blanton2006,abbas2007,Wu2024,pujol2017,paranjape2018,balaguera2024}. For a review of this topic, we refer the reader to \cite{wechsler&tinker2018}.

Galaxies inherit the secondary bias of their host halos, which influences the baryonic processes they undergo and, consequently, their growth and evolution. The strength of this inheritance depends on the galaxy type. Luminous Red Galaxies (LRGs) which quenched their star formation early, are primarily shaped by their halos’ assembly history rather than their current mass, making secondary halo bias a key driver of their properties. In contrast, Emission-Line Galaxies (ELGs) are actively forming stars, relying on the gas currently available in their halos, which is primarily determined by the halo’s present mass. As a result, ELGs are largely insensitive to secondary halo biases. We note that while a mass-based approach may captures the overall clustering of ELGs, they tend to be found in lower-density regions such as filaments and sheets \citep{gonzalez2020} so incorporating these environmental preference it is important for high-precision analyses or studies of small-scale galaxy evolution \citep{favole2016,gonzalez2020,Hadzhiyska2021,rocher2023,yu2024,yuan2025}.

\begin{figure*}[h]
    \centering
    \includegraphics[width=18cm]{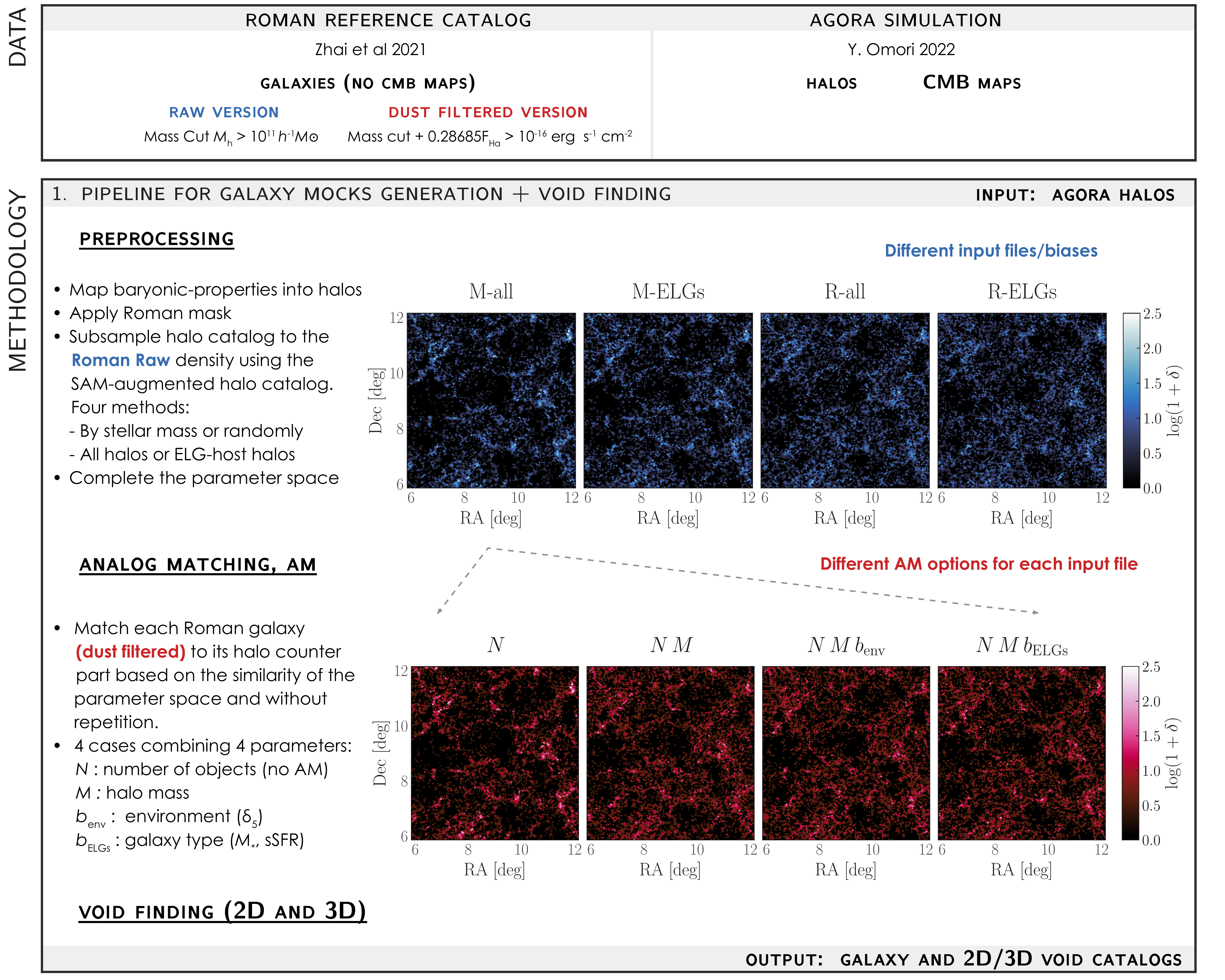} 
    \caption{Flowchart illustrating the datasets and methodology employed in this paper.}
    \label{fig:infography}
\end{figure*}

The subtle dependencies that define the galaxy–halo connection also affect void physics, since voids are traced indirectly through galaxies. Several studies have explored this interplay from different perspectives. HOD-based analyses and observations have shown that voids host distinct galaxy populations---fainter, bluer, and younger galaxies residing in halos with fewer satellites \citep{alfaro2020, alfaro2022,dominguez2023galaxies,conrado2024cavity}---and that void statistics are sensitive to variations in HOD parametrizations, including environmental effects in galaxy formation \citep{benson2001, tinker2006}. The work of \citet{nadathur2015_impact_mocks_in_voids} demonstrated that using unbiased dark matter particles as tracers---subsampled to match the mean galaxy number density, as was common in early analyses---fails to capture the effects of galaxy bias, leading to significant discrepancies in void counts and density profiles. At fixed sampling density, biased tracers produce far fewer small and intermediate voids, with differences comparable to those induced by changes in cosmological parameters.

Simulations have revealed that halo secondary properties, such as spin and concentration, also differ in voids: low-mass halos tend to have lower spin, more elongated shapes, and are more dynamically relaxed due to reduced merger activity, while mass-related quantities (like halo mass or peak velocity) remain largely unchanged \citep{balaguera2024}. 

Furthermore, there are studies that have assessed the extent to which galaxy voids trace the underlying dark matter field \citep{sutter2013}. Yet, no previous work has systematically investigated how different mock catalogs and modeling complexities impact both void statistics and Void-CMB cross-correlation studies. 

\subsection{Our method: analog matching}

Our approach to create the mock catalog, which we call \emph{analog matching}, does not fall into the aforementioned standard physical or empirical categories, although it shares some of their principles. In this work, we do not need to construct a mock catalog from scratch to match an observational dataset. We instead rely on the \emph{Roman} mock catalog by \cite{zhai2021clustering} as a reference and aim to generate an analogous mock for a simulation with self-consistent CMB maps, like \textsc{Agora}. The overall workflow of our methodology is summarized in Fig.~\ref{fig:infography}.

The core of the procedure is to identify an \textsc{Agora} halo counterpart for each \emph{Roman} mock galaxy using a nearest-neighbor search implemented with the Python package KDTree \citep{2020SciPy-NMeth}. A KDTree is a non-parametric data structure optimized for nearest-neighbor searches in $k$-dimensional spaces that can capture complex, non-monotonic correlations beyond simple rank-order models. Nearest-neighbor methods are common analytical tools in cosmology---for example, to extract higher-order statistics of the density field \citep{banerjee2021}---but have also been applied in mock catalog generation \citep{hearin2020generating,wechsler2022addgals}.

Initially, the \emph{Roman} reference mock catalog contains more than ten times the number of galaxies expected to be observed in the redshift range $1<z<2$. To reproduce realistic survey conditions, we adopt the \cite{Calzetti2000} dust model to scale the \ion{H}{$\alpha$} fluxes to the observed values and apply a flux cut to select only detectable galaxies. The resulting working dataset includes: a raw galaxy catalog, containing all sources\footnote{satisfying: $\max(\mathrm{H}\alpha, [\mathrm{O\,III}]) > 5\times10^{-18}\ \mathrm{erg\ s^{-1}\ cm^{-2}}$ (dust-free flux) and $H < 25$ assuming dust attenuation with $A_V = 1.6523$.} and a dust-filtered catalog, which represents the realistic observed sample after the flux cut is applied. In both cases we impose the halo mass threshold of $10^{11} h^{-1}\rm{M}_\odot$, to match the resolution of the \textsc{Agora} simulation, leading to negligible changes (of 1.32$\%$) in the $dN/dz$ for the dust-filtered catalog and the redshift range considered. 

Each \emph{Roman} galaxy is represented in a multidimensional parameter space by a vector, $\chi_{i,\text{REF}}$, whose components include halo mass, environmental indicators, and other galaxy properties such as stellar mass or specific star formation rate. An equivalent vector, $\chi_{i,\text{\textsc{Agora}}}$, is computed for the \textsc{Agora} halos and a KDTree is used to establish a correspondence between both. Matching is performed in redshift slices of 0.25~$h^{-1}$~Mpc.

Constructing these vectors requires some preprocessing. For the \emph{Roman} reference catalog, this is minimal: the $\chi_{i,\mathrm{REF}}$ components are obtained directly from the catalog except for the environmental parameter, which we compute identically to \textsc{Agora}. By contrast, the \textsc{Agora} halo catalogs require preprocessing to imprint Roman’s footprint and to extend $\chi_{i,\textsc{Agora}}$ beyond halo mass. This involves augmenting the halo catalog with baryonic information and computing the environmental parameter consistently with the \emph{Roman} catalog, a process that requires subsampling the halo dataset. Details follow.

\subsubsection*{{a) Halo catalog preprocessing: \\Parameter Computation and Subsampling.}}

We begin by coulping the \textsc{MultiDark Planck 2} halos embedded in \textsc{Agora} to the \textsc{UniverseMachine} galaxy formation model \citep{behroozi2019universemachine}, which predicts stellar mass ($\rm M_*$) and star formation rate (SFR) for each subhalo. From these, we assign stellar mass and the specific star formation rate ${\rm sSFR= SFR}/\rm{M}_*$ to each halo in \textsc{Agora}, and use the bimodal sSFR distribution in the $\rm{M}_*$–sSFR plane \citep{blanton2009physical} to distinguish between star-forming galaxies (including ELGs) and quenched galaxies (including LRGs), following the same approach as \texttt{Skyline} \citep{JB2023}. 

Then we mask the \textsc{Agora} simulation to match the \emph{Roman} survey footprint and subsample it to the expected \emph{Roman} raw galaxy number density, so that the number of \textsc{Agora} parent halos (subhalos) equals the number of central (satellite) galaxies in the \emph{Roman} raw mock catalog. The subsampling serves several purposes. First, by reducing the number of halos, it creates more manageable files which speed up the performance of the KDTrees. Second, it ensures a fair comparison between the \emph{Roman} reference catalog, which contains only galaxies, and \textsc{Agora}, which contains halos. Subsampling allows us to select a subset of halos representing potential observed galaxies and also ensures that environmental indicators are computed consistently between the datasets, since these metrics depend on the tracer bias when not derived from the underlying dark matter field. Later on, with the \emph{analog matching}  algorithm, we would further bias this selection towards something that resembles the dust-filtered \emph{Roman} mock catalog but we consider this step poses a more realistic and consistent baseline. Third, it allows us to explore how different subsampling strategies or tracer selections in the \textsc{Agora} simulation affect the performance of the \emph{analog matching} algorithm.

Using the baryonic properties mapped onto \textsc{Agora}, we subsample the halo catalogs in four ways: ranking halos by their assigned stellar mass or choosing them randomly, and in each case selecting either the full halo population (hosts of all-type galaxies) or only halos labeled as ELG hosts (see Figure \ref{fig:UM_vs_galacticus_prop}). 

Finally, once the subsample is done and to complete the parameter set, we account for the environmental effects by computing the mass-weighted density contrast, formally defined as follows:

\begin{equation}
\mathcal{\delta}^{(i)}_R = \frac{\sum_{i} M_h}{\left<M_h\right>}-1 
\end{equation}

For each tracer, we count the number of neighboring tracers within a sphere of radius $\mathcal{R}$, weighting them by their halo masses and normalizing by the mean halo mass in the corresponding redshift bin and inside the same volume. We choose $\mathcal{R} = 5~h^{-1}$~Mpc following previous studies \citep{balaguera2024,Wu2024} that claim that this radius effectively captures small-scale clustering. 

We weight by halo mass, $M_h$, to emphasize the different environments in which galaxies can live. Although the \emph{Roman} catalog is flux- and magnitude-limited, its raw version is denser than the dust-filtered sample and therefore provides a better approximation of the underlying matter distribution. To enable a consistent comparison, we compute the density contrast in the raw catalog and in the sub-sampled \textsc{Agora} files (for each of the four different biases). These calculations are performed for each central galaxy, after which satellites are assigned the density contrast of their central (or closest central). 

We need to take into account that the \emph{analog matching} algorithm is applied to the final, dust-filtered catalog rather than the raw one even though some properties such as the environmental parameter ($\delta_5$) %the density contrast, 
is computed from the raw sample. We use the dust-filtered version to ensure that the best-fitting analogs from the sub-sampled \textsc{Agora} simulation are reserved for the galaxies that will actually comprise our final sample. By doing this, we avoid wasting close neighbors on galaxies that would later be discarded by observational cuts.\\

\subsubsection*{{b) Analog Matching through KDTrees.}} After defining the parameter space, $\chi$, for both the sub-sampled \textsc{Agora} simulation and the \emph{Roman} dust-filtered reference catalog, we standardize each parameter to ensure that all the properties contribute equally to the matching. For each \textsc{Agora} halo attribute $\chi_{i,\rm \textsc{Agora}}$, (halo mass, environmental parameter, stellar mass or specific star formation rate), we subtract the mean of the corresponding parameter in the \emph{Roman} catalog and divide by its standard deviation:

\begin{equation}
   \mathrm{ \chi_{i, \textsc{Agora,stand}} = \frac{{}\chi_{i,\textsc{Agora}} - \overline{\chi_{i,\textsc{Ref}}}}{\sigma_{\chi_{i,\textsc{Ref}}}}}.
\end{equation}

Standardizing with respect to the \emph{Roman} catalog ensures that the matching is properly weighted toward the observed sample.

 An important consideration is that finding counterparts between two samples can produce duplicates; that is, a single \textsc{Agora} halo could be the closest match for multiple \emph{Roman} galaxies. To prevent this, once a halo has been assigned, it is removed from the pool of available neighbors. The first objects to be matched receive the closest counterparts, so the order of the pairing is important. Following the logic of SHAM models, we prioritize mass. We rank \emph{Roman} galaxies from most to least massive and assign nearest neighbors in the standardized parameter space. If a halo has already been used, the algorithm selects the next-closest match, ensuring a one-to-one correspondence while prioritizing the most massive, and thus most clustering-relevant, objects in the final catalog.

On this basis, we construct four families of mock catalogs, each reflecting a different level of specification. The names of these are: 

\begin{list}{}{}
	\item [\textbullet] {$N$ (number)}: This serves as a baseline method and no \emph{analog matching} is applied at this stage. The procedure is the same as the first subsampling of the \textsc{Agora} files (with four different biases), but in this case we directly match the number density of the dust-filtered catalog, without considering detailed halo properties.
	\item [\textbullet] {$N M$ (halo mass)}: This method uses the \emph{analog matching} technique with halo mass alone.
	\item [\textbullet] {$N M b_{\rm env}$ (halo mass and environment measures)}: This approach accounts for both halo mass and the local environment, quantified by $\delta_5$.
	\item [\textbullet] {$N M b_{\rm ELGs}$ (halo mass and galaxy-type indicators)}: This method integrates halo mass and a parameter indicative of galaxy type, such as the stellar mass ($\rm M_{*}$) and the specific star formation rate (sSFR). This is the most similar to a SAM approach.
\end{list}

Each mock family is generated for the four initial \textsc{Agora} subsampling variants using the baryonic information of the SAM-augmented halo catalog (by stellar mass or randomly, considering either all galaxy types or only ELGs) resulting in a total of sixteen mock catalogs: four mock types, each run for four slightly differently biased files (see Figures~\ref{fig:infography} for a schematic explanation and \ref{fig:UM_vs_galacticus_prop} for clarification purposes).

\subsubsection*{c) Tests} We validate the resulting catalogs by comparing them to the dust-filtered catalog across a number of one-point and two-point statistics. The former group includes galaxy and void redshift distributions, void size functions, and void minimum density distributions, while the latter includes the matter power spectrum and the void–galaxy cross-correlation function.

\subsubsection*{Strengths and limitations}
Our method does not aim to compete with the most sophisticated empirical frameworks. Instead, it provides a practical middle ground for building analog catalogs from existing ones. By incorporating multiple galaxy and halo properties simultaneously, it captures complex, multivariate relationships that one-dimensional methods such as SHAMs cannot resolve. At the same time, the use of KDTrees ensures that even large datasets can be processed efficiently, enabling fast generation of diverse mock realizations. 

Naturally, the method carries certain limitations that must be acknowledged and which we address and discuss next.

\begin{list}{-}{}
\item [\textbullet] {Parameter consistency: } One potential issue appears when- the physical properties for the matching are defined or computed differently across datasets. Some features, such as the local environment, intended as a proxy for the matter distribution surrounding each galaxy, are affected by sample biases. In the reference mock, the raw catalog contains more galaxies than the dust-filtered one, yet is still biased by selection cuts. To make a fair comparison, we subsample the \textsc{Agora} halo catalogs to the number density of the \emph{Roman} raw mock catalog, exploring multiple subsampling strategies to emulate these biases. If the bias still differ, the environmental parameters remain non comparable and the pairings become unreliable, leading to systematic deviations in one- and two-point statistics.
\bigskip
\item [\textbullet] {Spatial completeness and masking: } Observational systematics that affect the spatial distribution of the observed galaxies, such as survey edges or other completeness losses, are not included in the algorithm, as these are absent in the reference catalog. In our method, these effects must be applied as a preprocessing step on the initial \textsc{Agora} halo files. Since the \emph{analog matching} is designed to fix the number density---by finding a counterpart for every reference object uniformly across the footprint---, applying completeness variations is necessary beforehand, to preserve the desired number density while reproducing the spatial variations. If applied afterwards, objects would need to be removed lowering the desired number density.
\bigskip
\item [\textbullet] {Ranking order: } Finally, enforcing unique matches (so that each simulated object is assigned to only one reference object) means that the order of assignments can affect results, potentially leading to suboptimal pairings in crowded regions of the parameter space. Although alternative strategies could be explored, we prioritize a mass-based ranking inspired by SHAM methods, which has proven effective for our purposes.
\end{list}

\subsection{Cosmic voids as mock validation tools}

\begin{figure*}[h]
    \centering
    \includegraphics[width=17.5cm]{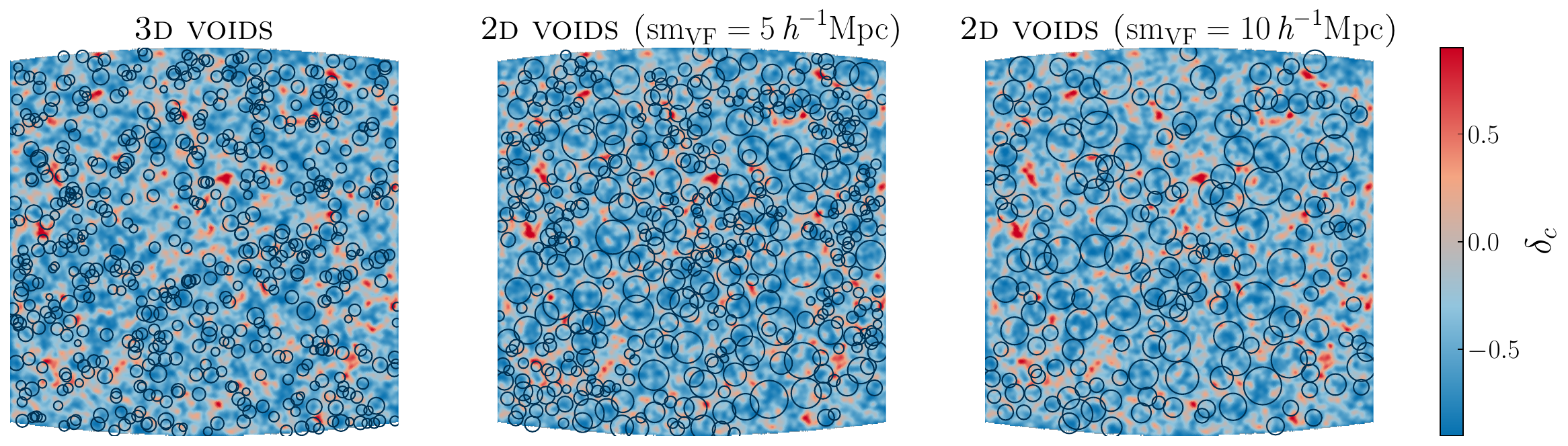} 
    \caption{Slice at $z = 1.012$ with thickness $\Delta z = 0.01$, illustrating 3D and 2D voids (identified  with smoothing scales of 5 and 10~$h^{-1}$~Mpc).  To maintain clarity, only voids centered at this exact redshift are plotted; otherwise, including the full slice width would result in 3D voids covering the map. Specifically, we identify 525 3D voids (restricted to this redshift), compared to 470 2D voids at $\rm sm_{VF} =  5$~$h^{-1}$~Mpc and 209 at $\rm sm_{VF} =  10$~$h^{-1}$~Mpc. Their average sizes are $24.91, 50.38,$ and $83.90$~$h^{-1}$~Mpc (corresponding to angular sizes of $0.42, 0.84$, and $1.48$~degrees), respectively.}
\label{fig:slice_with_voids}
\end{figure*}

To further probe the galaxy distribution beyond 2-point correlation functions, we analyze the statistical properties of cosmic voids in our mock catalogs. 

Given the diversity of void definitions and finders, there is no unique prescription for selecting a method. In practice one adopts the definition that maximizes the signal for the observable of interest. Projected (2D) voids, for instance, provide strong lensing signals \citep{cautun2018}, although 3D outnumber them by orders of magnitude and can also deliver competitive measurements \citep{Fang2019}. Since our aim is not to test any theoretical model but to quantify the impact of different mock catalogs on void statistics, we employ two pragmatic and widely used void finders---a 2D (density-based) and a 3D (geometrical) one---chosen for their relevance to forthcoming void–CMB cross-correlation studies.

\paragraph{\textbf{2D Voids}}

For identifying 2D voids we use the algorithm by \cite{sanchez20162d}. The code works by projecting galaxies into 2D slices and finding voids in the smoothed 2D galaxy density field of the slice. The void finding process involves the following steps:

{\it (i)} Divide the tracer sample into redshift slices of thickness $s$ in comoving distance. {\it (ii)} For each slice, a density contrast map is calculated by projecting tracers onto a HEALpix map\footnote{$N_{\textrm{side}} = 512$ representing an angular resolution of 0.1 deg and a physical resolution of 4~$h^{-1}$~Mpc at $z=1$ (6~$h^{-1}$~Mpc at $z=2$).} and smoothing the field with a Gaussian filter of comoving scale $\rm sm_{VF}$. {\it (iii)} Voids are identified by locating the most underdense pixels ($\delta < -\sigma_{\delta_c}$, \footnote{meaning the local density is lower than the cosmic mean by more than one standard deviation of the density fluctuations.}) as initial centers, and a circular region of radius $R_v$ is grown around each center until the mean density is reached. {\it (iv)} Identified void pixels are excluded from further consideration, and the process iterates until all pixels are assigned to voids. {\it (v)} Voids that extend beyond the survey boundaries by more than a defined fraction of their area are excluded. This is determined using a random point catalog by comparing the number of random points within each void to the expected count from the mean random-point density; those retaining less than a threshold fraction (e.g., 70 $\%$) are deemed incomplete and removed. {\it (vi)} Finally to improve line-of-sight positioning, the slicing process is repeated with shifted centers, and voids in neighboring slices with small angular separations are grouped into single structures. The final positions of these grouped voids are computed as the median positions of their constituent members.

The 2D void identification depends on several free parameters, which influence the resulting void catalogs and observables \citep[see e.g.,][]{vielzeuf2021desy1}. These include the Gaussian smoothing scale ($\rm sm_{VF}$) applied to the projected galaxy density field and the thickness of the tomographic slices. We adopt a smoothing parameter of $\rm sm_{VF}$ = 10~$h^{-1}$~Mpc and a slice thickness of $s \approx 70$ ~$h^{-1}$~Mpc although other configurations were tested\footnote{Most studies employing 2D void finders rely on photometric redshifts, where the photo-$z$ scatter forces the use of thick tomographic slices of $\sim$100~$h^{-1}$~Mpc  \citep{sanchez20162d} to ensure the detection of independent underdense structures. Considering that we work with spec-$z$s, thinner slices are feasible.}.

The outputs of the void finder are the position of the voids (ra, dec, $z_{obs}$); the transversal and radial sizes; the mean density contrast $\bar{\delta} (r<r_v) = \rho/\bar{\rho}-1$ where $\rho$ is the average density inside the void and $\bar{\rho}$ is the mean density of the corresponding redshift slice; and the central density contrast which corresponds to the density contrast evaluated at one quarter of the void radius $\delta_{\rm 1/4}\equiv \delta(r=0.25~r_v)$.

\paragraph{\textbf{3D Voids}}

For identifying 3D voids, we use \texttt{REVOLVER}\footnote{\url{https://github.com/seshnadathur/Revolver}} (REal-space VOid Locations from surVEy Reconstruction), which is based on a modified version of the \texttt{ZOBOV} algorithm \citep{neyrinck2008zobov}. The void identification process includes the following steps: \\
{\it (i)} The tracer density field is reconstructed using a Voronoi Tessellation Field Estimator (VTFE), which divides the survey volume into Voronoi cells---one per tracer---and assigns each cell a density equal to the inverse of its volume. {\it (ii)} Void centers are identified at local minima, and neighboring regions are merged into 'zones' following rising density gradients. Boundaries between voids are identified at saddle points, i.e, where the density stops increasing and begins to slope toward an adjacent minimum. These ridges delineate distinct density basins, each of which is treated as an independent void without further merging. {\it (iii)} Void centers are assigned as the circumcenter of the lowest-density tracer and its three lowest-density adjacent neighbors (although the volume-weighted barycenter is also provided for comparison purposes) \footnote{The circumcenter is the point equidistant from all chosen vertices, representing the geometric center of the underdense region, while the barycenter is the arithmetic mean of all points weighted by their cell volume.} {\it (v)} Finally, the effective void radius is determined as the radius of a sphere with a volume equal to the total volume of the constituent Voronoi cells.

Survey edges induce artificially large Voronoi volumes. The algorithm buffers the mask with fake mock tracers simulating overdense regions and discards those and nearby cells before the watershed step.

The output of \texttt{REVOLVER} includes: the void ID; the void center ($\mathbf{X}_v$) with both the circumcenter and barycenter; the effective spherical radius ($R_v$); the central density contrast ($\delta_{v,\rm{min}}$); the average density contrast ($\overline\delta_v$); the  density ratio, i.e, the ratio of the highest density at the void’s edge to the minimum density at the center; and the dimensionless parameter ($\lambda_\mathrm{v}$), a proxy for the gravitational potential, which distinguishes different void populations and plays a role in their lensing properties \citep[see e.g.][]{sheth&vdweygaert2004_vsf,raghunathan2020cmbk}. Negative $\lambda_\mathrm{v}$ indicates large underdense voids (void-in-voids), while positive values correspond to smaller overdense regions (void-in-clouds):

\begin{equation}
    \lambda_\mathrm{v} \equiv \overline{\delta}_\mathrm{v} \left( \frac{R_\mathrm{v}}{1\, h^{-1}\mathrm{Mpc}}\right)^{1.2} \, .
\end{equation}

\textsc{Zobov} operates with minimal assumptions and no free parameters which allows to topologically detect voids purely from the density field. 
 
\section{Results}
\label{sec:results}

\begin{figure*}[h]
    \centering
    \hspace*{-0.3cm} 
    \includegraphics[width=18.cm]{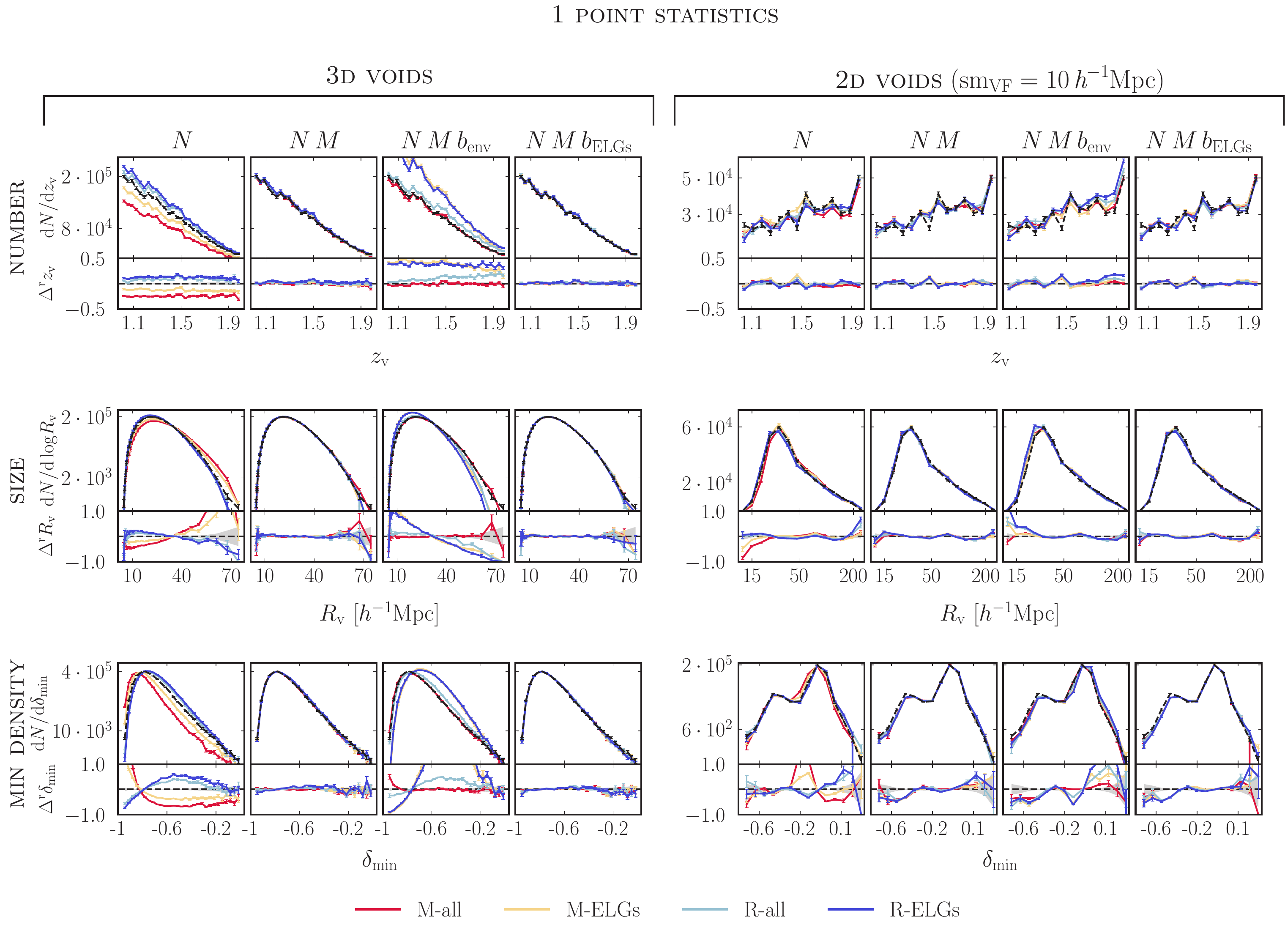}
    \caption{One-point statistic for the generated void catalogs. The rows correspond to the void redshift (top), void size (middle), and minimum density (bottom) distributions for the respective void samples. Each column represents a different mock family. Different colors denote a different tracer selection in the input files. The bottom sub-panels display the residuals, defined as the relative difference with respect to the \emph{Roman} reference catalog. Error bars are Poissonian.}
    \label{fig:one-point_main}
\end{figure*}

Here we present how different mock configurations impact the resulting galaxy clustering and void statistics. Each mock is designed to reproduce the \emph{Roman} reference catalog with a distinct level of accuracy.

\subsection{One-point statistics}
\label{sec:one_point_stats}

For galaxies, the one-point statistics essentially corresponds to the number density or redshift distribution ($dN/dz$). We ensure this matches the one of the reference catalog by construction, since each \emph{Roman} galaxy is assigned a corresponding analog in the \textsc{Agora} simulation. For voids, we analyze the one-point statistics given by the void redshift distribution, the void size function, and the minimum density distribution.

All plots in the paper follow a consistent layout to make comparisons easier: columns distinguish between mock families, while colors represent different tracer selections in the input files. Each subplot displays the residuals of the corresponding observable relative to the \emph{Roman} reference catalog.

The mock families are from left to right: simple subsampling ($N$), \emph{analog matching} based on halo mass alone ($NM$); halo mass plus environment ($NMb_{\rm env}$); and halo mass plus galaxy-type indicators ($NMb_{\rm ELGs}$). 
Within each family, the tracer selections are defined by how the \textsc{Agora} simulation is subsampled before applying the \emph{analog matching} algorithm. In particular, halos are selected either randomly (R) or ranking them by their assigned stellar mass (M), and in each case considering either the full halo population (-all) or only halos labeled as ELG hosts (-ELGs). For the $N$-mock family, we directly subsample the catalog to match the number density of the dust-filtered case using these four criteria  (M-all, M-ELGs, R-all, R-ELGs).

Figure \ref{fig:one-point_main} shows the one-point statistics for 2D and 3D voids. The largest discrepancies with respect to the reference case are observed for the $N$ and $NMb_{\rm env}$ mock catalogs. 

Within the $N$-family of subsampled catalogs, the mass-selected sample (M-all) shows the biggest deviation. This is expected as ELGs typically inhabit lower-density regions whereas selecting the most massive tracers emphasizes clusters and filaments in the cosmic web, resulting in fewer but larger and deeper voids. When selecting tracers by both stellar mass and ELG-type (M-ELGs), the distribution becomes closer to the reference case. Mock galaxies are now present in lower-density regions as well, which divides previously-large voids into smaller ones. Randomly selecting tracers (R-all and R-ELGs) further amplifies this effect, creating even more fragmented and shallower voids as tracers populate all types of environments in this case. These random-selected samples, either for the case of all galaxy types or only ELGs, are very similar. Overall, none of the $N$-type mock catalog selections fully reproduce the trend observed in the \emph{Roman} reference catalog, although the R-ELG and R-all cases show comparatively good agreement.

For the $NMb_{\rm env}$ family, in the third column of each block of plots, the observed discrepancies stem from the algorithm itself, highlighting one of the aforementioned limitations of the \emph{analog matching} method. This technique requires that the quantities compared between the reference mock and the target simulation represent the same physical property. The $b_{\rm env}$ parameter, which quantifies the environment through the density contrast in spheres of 5~$h^{-1}$~Mpc, is heavily influenced by the type of tracer used. Ideally, we would compute b$_{\rm env}$ directly from the dark matter field for both datasets, but as mentioned, the raw mock catalog includes only galaxies with no information about the full halo population or the underlying matter field. Among the subsampled \textsc{Agora} halos, designed to replicate the bias of the reference mock catalog, the mass-selected sample (M-all) reproduces its properties most closely (see Appendix \ref{fig:UM_vs_galacticus_prop} for details).

For the families based on mass ($NM$), or the more complex implementation including mass and $b_{\rm ELGs}$ ($NMb_{\rm ELGs}$), we find excellent agreement with respect to the reference case for all tracer selections. Neither of the two options shows a clear advantage, suggesting that halo mass alone is sufficient to reproduce void statistics coming from ELGs. The largest deviations occur only at the extreme values of $\delta_{\rm min}$, for very large or very shallow voids, but the overall trends remain consistent. 

All the effects discussed can be seen more clearly for 3D voids. 2D voids under the standard smoothing scale of $10~h^{-1}$~Mpc wash out much of the catalog-to-catalog variation. If we reduce the smoothing to $5~h^{-1}$~Mpc, the differences start to be apparent although still more subtle than for the 3D case (see Appendix Fig.~\ref{fig:one_and_two_point_for_2Dvoids_with_sm5}).

\begin{figure*}[h]
    \centering
    \includegraphics[width=14cm]{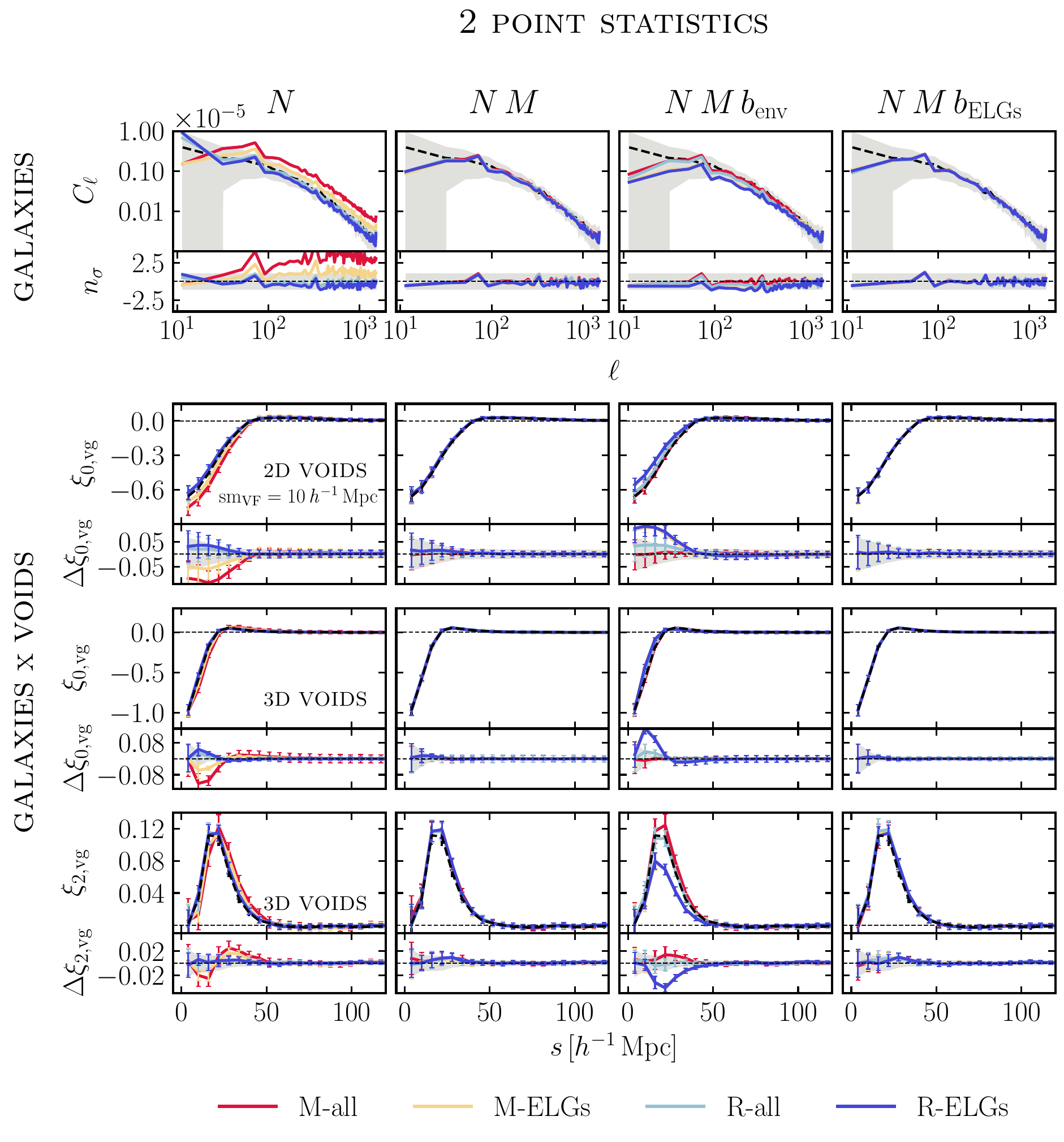} 
    \caption{Two-point statistics including the galaxy power spectrum (top panel) and the galaxy-void cross-correlation function (bottom panel). For the latter, the monopole of both 2D and 3D voids and the quadrupole of the 3D voids are included. Each column corresponds to a different mock family, and each color represents a different tracer selection. The residuals in the bottom sub-panels show the relative difference compared to the \emph{Roman} reference catalog. Error bars are derived from jackknife estimates.}
    \label{fig:two-point_main}
\end{figure*}

\subsection{Two-point statistics: methods}
\label{sec:two_point_stats}

We extend our analysis to two-point statistics by examining the galaxy angular power spectrum and the void-galaxy cross-correlation function for both 2D and 3D voids. We first define these quantities, describe how they are computed, and then present and discuss the results.

The \textbf{angular power spectrum}, $C_\ell$, quantifies the clustering of matter, in 2D projected count maps, by measuring the variance of density fluctuations as a function of angular scale $\ell \sim \pi/\theta$. Higher $C_\ell$ amplitudes point to stronger anisotropies, corresponding to either overdense regions (clusters) and/or underdense regions (voids). Lower amplitudes indicate relatively uniform regions with little structure. We compute the $C_{\ell s}$  using \texttt{NaMaster}   \citep{alonso2019namaster}, and the error bars are estimated from the standard expression \citep{knox1995}:
\begin{equation}
\sigma_{C_\ell} = \sqrt{\frac{2}{(2\ell + 1) f_{\rm sky}}} \,\left(C_\ell + N_\ell \right),
\end{equation}
where $f_{\rm sky}$ is the observed sky fraction, $C_\ell$ the angular power spectrum, and $N_\ell$ is the shot noise power spectrum, equal to $N_\ell=1/\bar{n}_g$, with $\bar{n}_g$ the average galaxy angular number density.

The \textbf{Void-Galaxy Cross-Correlation Function} (VGCF), $\xi(r,\mu)$, measures the excess probability of finding a galaxy at distance $r$ (from a void center) and angle $\mu = \cos\theta$ (relative to the line of sight toward the void center), compared to a random and unclustered galaxy distribution. Under the cosmological principle, which states that the Universe is statistically homogeneous and isotropic on large scales, stacked voids should appear spherical when averaged, implying that $\xi(r,\mu)$ would depend only on $r$ \citep{lavaux&wandelt2012_stacking_vgcf_aprsd}. However, the observed VGCF reveal anisotropies due to two effects: dynamical distortions from galaxy peculiar velocities along the line of sight (redshift-space distortions, RSD), and geometrical distortions from adopting an incorrect cosmological model when converting redshifts and angles into distances \citep[Alcock–Paczynski effect, AP,][]{alcock_1979}. By modeling these distortions, the VGCF has become one of the standard tools in void cosmology for constraining cosmological parameters \citep{alcock_1979,ryden_1995, sutter2012_vgcf_ap,mao2017_vgcf_ap, nadathur2020_vgcf_aprsd, hamaus2020_vgcf_aprsd,hamaus_2022_vgcf_aprsd, woodfinden2022_vgcf_aprsd,correa2022_vgcf_aprsd, radinovic2023_vgcf_aprsd,verza2024_vgcf_and_vsf_aprsd,fraser2025_vgcf_aprsd,degni2025_vgcf_aprsd}. We measure the VGCF using the Landy-Szalay estimator \citep{landy_szalay}:  

\begin{equation}
\xi(r,\mu) \;=\; 
\frac{ 
\langle D_v D_g \rangle - \langle D_v R_g \rangle - \langle R_v D_g \rangle + \langle R_v R_g \rangle 
}{ 
\langle R_v R_g \rangle },
\end{equation}

where $D$ and $R$ refer to data and random catalogs, respectively, and angled brackets indicate normalized pair counts between void centers ($v$) and galaxies ($g$). The estimator works by comparing the observed void-galaxy pair counts to those from synthetic random catalogs of voids and galaxies that share the same survey volume and selection function but contain no intrinsic clustering. In this work, we compute pair counts in 25 equally spaced bins in both $r$ and $\mu$ using \texttt{Corrfunc} from \citet{Corrfunc}. 

Once $\xi(r,\mu)$ is computed, we separate the isotropic and anisotropic contributions by expanding it into multipoles with Legendre polynomials $P_\ell(\mu)$:  

\begin{equation}
\xi_\ell(r) \;=\; \frac{2\ell + 1}{2} \int_{-1}^{1} \xi(r,\mu)\, P_\ell(\mu)\, d\mu.
\end{equation}

The \textbf{monopole}, $\xi_0(r)$, describes the average radial density profile of galaxies around voids. It is negative inside the void (fewer galaxies than average) and positive in the surrounding shell. 

The \textbf{quadrupole}, $\xi_2(r)$, captures the anisotropic distortions induced by RSD and AP effects. A null quadrupole indicates spherical symmetry while a negative or positive signal reflect deviations from sphericity. In particular, a positive quadrupole reflects elongation along the line of sight, typically from galaxy outflows, whereas a negative quadrupole indicates infall and a prolate shape. In our case, the small negative bump observed in $\xi_2(r)$ in Figure \ref{fig:two-point_main}, suggests the presence of compensation walls, where galaxies move from outside toward the void's edges.

Uncertainties are estimated via a jackknife resampling method. The galaxy and void samples are partitioned into $N=70$\footnote{We adopt this number to provide a sufficient number of jackknife realizations for a stable covariance estimate while maintaining regions large enough to capture the spatial correlations of voids and galaxies.} spatial regions using the $k$-means clustering algorithm (\texttt{scikit-learn}; \citealt{scikit-learn}). For each realization, one region is omitted, and the correlation function is recomputed with the remaining data.  The covariance matrix is then computed as:

\begin{equation}
C_{ij} = \frac{N-1}{N} \sum_{k=1}^{N} \left[ \xi^{(k)}(r_i) - \bar{\xi}(r_i) \right] \left[ \xi^{(k)}(r_j) - \bar{\xi}(r_j) \right],
\end{equation}

where $\bar{\xi}(r)$ is the mean of the jackknife realizations. The diagonal elements of $C_{ij}$ provide the variance and hence the error bars on the measurements.

When displaying the results, the plots are organized in the same way as for the one-point statistics. Every column represents one of the mock families ($N$:number; $NM$: number and mass; $NMb_{env}$: number, mass and environment; $NMb_{ELGs}$: number, mass and type of galaxy). Within each column, colors distinguish different tracer selections: M-all, M-ELGs (mass-selected), R-all, R-ELGs (random-selected).

\subsection{Two-point statistics: results}
\label{sec:void_gal_results}

The results of the two-point statistics are complementary to and consistent with the one-point statistics. The dispersion is larger for the $N$ and $NMb_{\rm env}$ families, whereas $NM$ and $NMb_{E\rm LGs}$ show good agreement with the \emph{Roman} reference catalog. The strength of the galaxy clustering, measured by the angular power spectrum ($C_\ell$), is directly related to the galaxy bias. This is clearly seen for the $N$-family, in the first column, where the different tracer selections produce a clear progression from most to least clustered with respect to the reference catalog. Mass-selected galaxies (M-all) are the most clustered, and this leads to the strongest density fluctuations: high densities in knots and sheets and very low densities elsewhere. In contrast, randomly selected galaxies (R-all) are distributed more evenly throughout the cosmic web, leading to a more uniform density field and weaker clustering. Selecting ELGs slightly reduce clustering in either case (M-ELGs or R-ELGs), smoothing out the overall density variations, as they preferentially occupy lower-density regions.

The VGCF monopole, $\xi_0(r)$, representing the average radial density profile, reflects the same clustering trends observed in the $C_\ell$ power spectra. The deepest voids are found in the M-all sample and become progressively shallower for the M-ELGs, R-all, and R-ELGs samples. The zero-crossing point, which is linked to the average void radius, shows that voids are larger in mass-selected samples, in agreement with the trends seen in the one-point statistics. This progression highlights how galaxy bias shapes void morphology: stronger clustering produces steeper density profiles and larger voids, while weaker clustering leads to smoother profiles.

The quadrupole, $\xi_2(r)$, exhibits a similar amplitude across the different tracer selections but is shifted in radius, aligning with the average void size indicated by the monopole.

The $NMb_{\rm env}$ family, as previously shown, only yields meaningful analogs when the galaxy bias in the \textsc{Agora} input files is similar to that of the reference catalog. If the bias differs, the same value of $b_{\rm env}$ can correspond to very different underlying environments, making the pairing unreliable. For instance, if the raw reference catalog is formed by a mass-selected sample of all type galaxies (M-all), as the results suggest, a low value of $b_{env}$ may correspond to a moderately dense filament while for a random-selected sample (R-all), the same value of $b_{\rm env}$ could be found in a void, since the underlying dark matter distribution is sampled more uniformly in this latter case.

Still within the $NMb_{\rm env}$ family, the ELG samples (either randomly or mass-selected) show a different behavior with respect the reference catalog, and a very similar behavior between them. Their effective bias seems to be lower than that of the reference catalog, the monopole is shallower for 2D and 3D voids, and the quadrupole is less distorted suggesting shallower and less elongated voids in redshift space.

Another important takeaway, that we can see in this mock family, is that similar galaxy clustering (i.e. similar $C_\ell$ power spectrum) does not guarantee similar void statistics. For instance, while the R-all sample exhibits a similar power spectrum to that of the M-all sample (and consequently to the reference case), its 3D voids differ significantly in size and density profile. This finding highlights that void statistics provide independent and sensitive information about how galaxies populate the cosmic web, beyond what is captured by the matter power spectrum, offering a powerful test for analyzing the accuracy of mock catalogs. 

For the $NM$ and $NMb_{\rm ELGs}$ families, all galaxy bias selections give consistent one- and two-point statistics, showing that the \emph{analog matching} method can effectively remove the initial bias in the input files when the same physical parameters are matched for both \textsc{Agora} and the reference catalog. The monopole and quadrupole are also very similar and fall within the error bars, making these the best match to the \emph{Roman} reference catalog.

It is important to note that none of the methods perfectly reproduces the $C_\ell$ at low $\ell$. This regime is dominated by cosmic variance, as reflected in the uncertainties. Cosmic variance arises because we observe only a single universe, and on very large scales there are few independent modes of density fluctuations to average over. This effect is further amplified when the survey does not cover the full sky, and measurements on these scales carry large statistical uncertainties. 

We nevertheless consider that this lack of agreement at low $\ell$ is not a major concern for void studies since their key properties, such as the size distribution and density profiles, are primarily determined by smaller scales (higher multipoles). Our most robust constraints are for scales below $\ell \sim 30 $ ($\sim$ 5.7$^\circ$), which corresponds to the largest 2D voids; with 3D voids being typically of smaller size. \footnote{3D voids span $0.11^\circ$–$1.31^\circ$ ($4$–$81\,h^{-1}\,\mathrm{Mpc}$, mean $22\,h^{-1}\,\mathrm{Mpc}$) in the redshift range $1<z<2$. For 2D voids with $5\,h^{-1}\,\mathrm{Mpc}$ smoothing, sizes range $0.09^\circ$–$4.97^\circ$ ($5$–$264\,h^{-1}\,\mathrm{Mpc}$, mean $33\,h^{-1}\,\mathrm{Mpc}$), and with $10\,h^{-1}\,\mathrm{Mpc}$ smoothing, $0.19^\circ$–$5.93^\circ$ ($11$–$264\,h^{-1}\,\mathrm{Mpc}$, mean $50\,h^{-1}\,\mathrm{Mpc}$).}

\section{Discussion \& Conclusions}
\label{sec:conclusions}

The correlation between cosmic voids and the CMB---most notably the Void $\times$ CMB lensing signal---has been firmly established observationally, with reported significances of 3-17~$\sigma$ \citep{cai2017cmbk_isw, chantavat2016cmbk, raghunathan2020cmbk, vielzeuf2021DESY1cmbk, hang2021cmbk_isw, kovacs2022DESY3cmbk, camachociurana2024cmbk, demirbozan2024cmbk, sartori2024cmbk}. However, several studies have also identified moderate ($\lesssim3\sigma$) tensions with respect to the $\Lambda$CDM predictions \citep{vielzeuf2021DESY1cmbk, hang2021cmbk_isw, kovacs2022DESY3cmbk, camachociurana2024cmbk}, the origin of which remains unclear. 

Properly interpreting these and future findings requires a careful distinction of methodological artifacts from genuine physical signals. Until the influence of analysis choices---such as void identification or mock catalog construction---is fully understood, observed tensions cannot be confidently attributed to physics beyond the standard cosmological model.

Motivated by this, and in preparation for the Nancy Grace $\emph{Roman}$ Space Telescope, we have developed an empirical framework designed to produce a set of $\emph{Roman}$-like mock catalogs of varying degrees of realism ranging from basic dark matter subsamples---known to inadequately reproduce galaxy and void statistics \citep{nadathur2015_impact_mocks_in_voids}---to survey-specific realizations that accurately recover the clustering and bias properties of the \emph{Roman} ELGs sample. These serve as a useful benchmark to investigate how catalog construction impacts different cosmological observables. 

Our algorithm, called \emph{analog matching}, reproduces existing galaxy mock catalogs in a simulation of interest. We employ the \textsc{Agora} simulation since it provides a full-sky halo lightcone alongside coherently modeled CMB secondary anisotropy maps. Within this framework, each galaxy from the reference \emph{Roman} mock catalog is associated with a halo in \textsc{Agora} through a nearest-neighbor search in a flexible parameter space. This space is highly customizable and may be as comprehensive as the user desires, including---among other parameters---halo mass, environmental indicators, and, when available, galaxy properties such as stellar mass and sSFR mapped to halos via precomputed SAMs. Galaxies are processed in decreasing order of mass so that massive objects are paired with their most suitable analogs preserving the high-mass end of the distribution.

We validate the resulting mock catalogs by comparing their one- and two-point galaxy and void statistics against those of the reference \emph{Roman} catalog:

\begin{itemize}
    \item[\textbullet] Our results confirm that both sets of statistics are highly sensitive to the details of the mock-construction process, consistent with previous studies \citep{berlind&weinberg2002_impact_mocks_in_voids, tinker2006_impact_mocks_in_voids, sutter2013_impact_mocks_in_voids, sutter2014_impact_mocks_in_voids, nadathur2015_impact_mocks_in_voids, contarini2019_impact_mocks_in_voids, salcedo2025_impact_mocks_in_voids}. 
    \smallskip
    \item[\textbullet] Importantly, we find that matching based on halo mass alone (or halo mass and galaxy type) is sufficient to reproduce the clustering and void statistics of \emph{Roman} ELGs, whereas including environmental information can degrade the algorithm's performance when the underlying dark matter field for both the simulation and the reference catalog is not available. 
    \smallskip
    \item[\textbullet] Throughout this paper, we find that void statistics provide a powerful diagnostic for assessing mock catalog accuracy beyond traditional clustering measures, with 3D voids generally showing stronger dependence than 2D voids. 
\end{itemize}

Looking ahead, it will be valuable to test how this methodology performs for tracers that exhibit stronger assembly bias, such as LRGs, for which halo mass alone is an insufficient predictor of clustering. Moreover, in forthcoming N-body simulations that include CMB companions and multiple cosmological realizations but lack baryonic information (e.g., $\rm M_{*}$ and sSFR), our method will require the reference catalogs to provide access to the underlying dark matter field. This will enable the computation of secondary halo bias indicators---such as tidal anisotropy, local matter density or cosmic-web classification \citep{balaguera2024}---to identify the halos where galaxies are most likely to be found, without assuming a direct link to galaxy bias.

In conclusion, rather than compete with the state-of-the-art mock-generation pipelines, our goal is to provide a validated suite of $\emph{Roman}$-like mock catalogs of varying accuracy that can serve as a controlled testbed for void–CMB correlation studies and related analyses. In a companion paper, we use the data products generated here to assess the impact of mock catalog construction on the measured Void$\times$CMB lensing signal.

\section*{Data availability}
The main products from this work, including mock galaxy catalogs, void catalogs, and analysis codes, will be made publicly available after acceptance of the paper for publication. We are available for consultation about the results or our methodology.

\section*{Acknowledgements}
This paper made use of the IAC HTCondor facility
(\url{http://research.cs.wisc.edu/htcondor/}), partly financed by the Ministry of Economy and Competitiveness with FEDER funds, code IACA13-3E-2493. MP wishes to acknowledge the contribution of the IAC High-Performance Computing support team and hardware facilities to the results of this research, specially to Ángel de Vicente. 

MP acknowledges Giulio Fabbian for useful discussions during the early stages of this work, as well as support from the pre-doctoral program at the Center for Computational Astrophysics, Flatiron Institute. Research at the Flatiron Institute is supported by the Simons Foundation. MP acknowledges support from the Agencia Estatal de Investigación del Ministerio de Ciencia en Innovación (AEIMICIN) and the European Social Fund (ESF+) under grant PRE2021-098156. MP and C.H.-M acknowledge the support of the Spanish Ministry of Science and Innovation under the grants PID2021-126616NB-I00 and "DarkMaps" PID2022-142142NB-I00, and from the European Union through the grant "UNDARK" of the Widening participation and spreading excellence program (project number 101159929). 

The Large-Scale Structure (LSS) research group at Konkoly Observatory has been supported by a \emph{Lend\"ulet} excellence grant by the Hungarian Academy of Sciences (MTA). This project has received funding from the European Union’s Horizon Europe research and innovation programme under the Marie Skłodowska-Curie grant agreement number 101130774. Funding for this project was also available in part through the Hungarian National Research, Development and Innovation Office (NKFIH, grant OTKA NN147550). 

AP acknowledges support from the European Research Council (ERC) under the European Union’s Horizon programme (COSMOBEST ERC funded project, grant agreement 101078174), as well as support from the French government under the France 2030 investment plan, as part of the Initiative d’Excellence d’Aix-Marseille Université - A*MIDEX AMX-22-CEI-03 as well as support from the Center for Computational Astrophysics, the Flatiron Institute and the Simons Foundation for early stages of this work.

%\newpage

\bibliography{bibliography}
\bibliographystyle{aasjournal}

\appendix

\section{\textsc{Agora} simulation details}
\label{appendix:agora_details}

The \textsc{Agora} lightcone is built by tiling the 1000 $h^{-1}$ Mpc simulation box to fill the required volume. The periodic boundary conditions ensures a continuous matter distribution across box edges. The observer is placed at the center, establishing the reference frame and from this point, spherical shells of 25~$h^{-1}$~Mpc width are extracted radially, with halos and dark matter particles from the closest simulation snapshot projected onto a HEALPix grid at $N_{\rm side} = 8192$ ($\sim 0.43$~arcmin resolution). To suppress artificial repetition along the line of sight, the shells are randomly rotated every 1000~$h^{-1}$~Mpc (one full box length). These rotations are applied at a scale significantly larger than the correlation length of the density fluctuations and consistently across all fields (halo catalogs, density, and velocity) to ensure accurate cross-correlations without inducing unphysical discontinuities.

The CMB $\kappa$ map is generated via multi-plane ray tracing. This method tracks light rays as they traverse successive spherical shells of the matter distribution, capturing cumulative and higher-order gravitational effects beyond the standard Born approximation\footnote{In the Born approximation, light rays are assumed to propagate along unperturbed, straight paths, and the total lensing distortion is computed as a single integral of the matter potential along that line of sight.}.
The procedure divides the redshift range into two regimes:

{\it (i)} For $z < 8.6$, the matter in each shell is converted into a 3D gravitational potential $\vec{\Phi}$ and then projected along the line of sight to define the 2D lensing potential:
\begin{equation}
 \phi (\vec{\theta}) = \frac{2}{c^2} \int_0^{\chi_{\rm LS}} d\chi \frac{\chi_{\rm LS}-\chi}{\chi_{\rm LS}\chi} \vec{\Phi} (\chi,\vec{\theta}),
\label{eq:lenseq1}
\end{equation}
$\chi$ denotes comoving distances, and $\chi_{\rm LS}$  is the comoving distance to the CMB last scattering surface at $z\simeq 1100$. From this quantity two fields are derived:
\begin{list}{}{}
    \item [\textbullet] The \textbf{deflection field} is defined by the first derivatives of the lensing potential $\vec{\alpha} (\vnh ) = \nabla_\Omega \phi$ and represents the angular displacement of the light rays.
    \item [\textbullet] The \textbf{tidal field} is defined by the second derivatives  $\vec{U_{ij}} = \nabla_{\Omega,i}\nabla_{\Omega,j}\,\phi(\vec{\vnh})$ and maps the local deformation of a light bundle by describing how gravity magnifies (convergence), stretches (shear), and twists (rotation) the light as it passes through each shell. 
\end{list}
The operator $\nabla_\Omega$ denotes the angular gradient on the unit sphere, defined as 
$\nabla_\Omega = \hat{\theta}\,\frac{\partial}{\partial \theta} + \frac{\hat{\varphi}}{\sin\theta}\,\frac{\partial}{\partial \varphi}$. Both the deflection and tidal fields are computed on a HEALPix \citep{healpix}\footnote{HEALPix's URL site \url{http://healpix.sf.net}} grid with $N_{\rm side} = 16384$ assuming a local thin-lens (Born-like) approximation for each shell. 

Using \textsc{GRayTrix} \citep{shirasaki2015probing}, light rays are traced backward from $z=0$ to $z = 8.6$  through successive lens planes of 25 $h^{-1}$ Mpc thickness. At each plane $j$, the deflection field $\vec{\alpha}$ shifts the ray's direction and the tidal matrix $\mathbf{U}$ is evaluated at the new coordinates to quantify the incremental distortion of the light bundle. These distortions are iteratively accumulated into the magnification matrix $\mathbf{A}$, updated at every shell via a recurrence relation that combines $\mathbf{U}_j$ with the previous $\mathbf{A}$ values, weighted by comoving angular-diameter distances $r(\chi)$. The final matrix $\mathbf{A}$ encodes the total convergence, shear, and rotation along each line of sight, from which the CMB lensing convergence is derived as:

\begin{equation}
\kappa^{\rm low\text{-}z}(\hat{\mathbf{n}}) = 1 - \frac{1}{2}\,\mathrm{Tr}\!\left(\mathbf{A}\right)
\end{equation}
The map is slightly smoothed to $N_{\rm side} = 8192$ to optimize computational efficiency but still retain small-scale accuracy. 

For $z > 8.6$, since the density field remains approximately linear and Gaussian, a Gaussian realization of the lensing potential is added, based on the theoretical power spectrum. 
\begin{equation}
\kappa^{\rm high\text{-}z}(\hat{\mathbf{n}}) 
= -\tfrac{1}{2}\,\nabla^2_{\Omega}\,\phi(\hat{\mathbf{n}})
\;\;\Longleftrightarrow\;\;
\kappa^{\rm high\text{-}z}_{\ell m} 
= \tfrac{1}{2}\,\ell(\ell+1)\,\phi_{\ell m} \, .
\end{equation}

The final map will be the sum of both contributions:
\begin{equation}
\kappa^{\rm tot}(\hat{\mathbf{n}}) = \kappa^{\rm low\text{-}z}(\hat{\mathbf{n}}) + \kappa^{\rm high\text{-}z}(\hat{\mathbf{n}})
\end{equation}

All the \textsc{Agora} CMB maps, including CMB $\kappa$, are calibrated and validated against observational data from \textit{Planck} \citep{planck2006overview}, SPT-SZ \citep{carlstrom2011}, and SPTpol \citep{bleem2012}, achieving sub-arcminute resolution and accurate statistics across $50 \lesssim \ell \lesssim 3000$. The LSS products (galaxy clustering and weak lensing) are similarly validated over $30 \lesssim \ell \lesssim 3000$ verifying that the simulation outputs are suitable for realistic cosmological studies.

Our analysis relies on the originally publicly released \textsc{Agora} products. While these are no longer publicly available at the time of submission of this work, the project webpage\footnote{\url{https://yomori.github.io/Agora/index.html}} indicates that updated versions are shared with collaborators.

\begin{table*}[]
\begin{adjustbox}{width=\textwidth}
\begin{tabular}{c|cccc|cccc|ccc}
\hline
\multirow{2}{*}{\textbf{\begin{tabular}[c]{@{}c@{}}Simulation\\ \small{Year}\end{tabular}}}                                    
& \multicolumn{4}{c|}{\textbf{\begin{tabular}[c]{@{}c@{}}General N-body \\ parameters\end{tabular}}} 
& \multicolumn{4}{c|}{\textbf{\begin{tabular}[c]{@{}c@{}}Cosmology \\ parameters\end{tabular}}} 
& \multicolumn{3}{c}{\textbf{\begin{tabular}[c]{@{}c@{}}\texttt{ROCKSTAR} \\ parameters\end{tabular}}}                                                                                                                            \\ \cline{2-12} 
& \textbf{\begin{tabular}[c]{@{}c@{}}Size \\ ($h^{-1}$~Mpc)\end{tabular}} 
& \textbf{Particles} 
& \textbf{\begin{tabular}[c]{@{}c@{}}res m$_p$ \\ ($h^{-1}$~M$_{\odot}$)\end{tabular}} 
& \textbf{\begin{tabular}[c]{@{}c@{}}log M$_h$ \\ ($h^{-1}$~M$_{\odot}$)\end{tabular}} 
& \textbf{$h$} & \textbf{$\Omega_m$} & \textbf{$\sigma_8$} & \textbf{$n_s$} 
& \textbf{\begin{tabular}[c]{@{}c@{}}Linking \\ length ($h^{-1}$~Mpc)\end{tabular}} 
& \textbf{\begin{tabular}[c]{@{}c@{}}Min number of\\ particles per halo\end{tabular}} 
& \textbf{\begin{tabular}[c]{@{}c@{}}Force \\ resolution\\ ($h^{-1}$~kpc)\end{tabular}} \\ \hline

\begin{tabular}[c]{@{}c@{}}\textbf{UNIT (Roman)}\\ \small{2016 (2021)}\end{tabular}  
& 1000  & $4096³$ & $1.2 \cdot 10^{9}$ & 8.18  
& 0.6774 & 0.3089 & 0.8147 & 0.9667  
& 0.2 L$_{box}$/N$_c$ & 20 & 6 \\

\begin{tabular}[c]{@{}c@{}}\textbf{MDPL2 (Agora)}\\ \small{2017 (2022)}\end{tabular} 
& 1000  & $3840³$ & $1.5 \cdot 10^{9}$ & 11.01  
& 0.6777 & 0.3071 & 0.8288 (0.818) & 0.96  
& 0.2 & $\geq$250 & \begin{tabular}[c]{@{}c@{}}13 (high-z) \\ 5 (low-z)\end{tabular} \\ \hline

\end{tabular}
\end{adjustbox}
\caption{Overview of UNIT and MDPL2, which serve as the basis for the \emph{Roman} reference catalog and the \textsc{Agora} simulation. We compare general N-body parameters, adopted cosmology, and \texttt{ROCKSTAR} halo finder settings.}
\label{tab:comparison_sims}
\end{table*}

\section{Comparison between UNIT and MDPL2}
\label{appendix:unit_vs_mdpl2}

Table \ref{tab:comparison_sims} compares the source simulations behind our datasets, UNIT for the \emph{Roman} catalog and MDPL2 for the \textsc{Agora} simulation, across three categories: (i) general N-body parameters, (ii) cosmological parameters, and (iii) the \texttt{ROCKSTAR} halo finder configuration \citep{behroozi2012rockstar_halos}, with merger histories computed using \texttt{CONSISTENT TREES}\footnote{\url{https://bitbucket.org/pbehroozi/consistent-trees/src/main/}} in both cases. 
%\chm{[CHM] Hmm, not sure if it is worth spending a few words explaining what \texttt{CONSISTENT TREES} is...}

\begin{list}{}{}
    \item [\textbullet] \textbf{General N-body parameters.} The \emph{Roman} reference catalog from UNIT has a higher resolution: for the same $1\,h^{-1}{\rm Gpc}$ box, it uses $4096^3$ particles compared to the $3840^{3}$ of MDPL2. This leads to a finer particle mass resolution of $1.2 \times 10^9~h^{-1}~$M$_{\odot}$ for UNIT, versus $1.5\times 10^9~h^{-1}~$M$_{\odot}$ for MDPL2. UNIT can resolve haloes down to $\log M_h \sim 8.2$, while MDPL2 reaches only $\log M_h \sim 11.0$, and this is due to the adoption of different particle thresholds at defining haloes (see below). 

    \item [\textbullet] \textbf{Cosmological parameters.} Both simulations adopt parameters consistent with \citep{ade2016planck}, with only minor differences. Notably, while MDPL2 uses $\sigma_8 = 0.8288$, the \textsc{Agora} products are generated with $\sigma_8 = 0.818$ to better match the amplitude of the matter power spectrum.

    \item [\textbullet]\textbf{\texttt{ROCKSTAR} parameters.} 
\begin{enumerate}
	\item \textit{Linking length:} defines the maximum distance at which particles are considered part of the same halo, expressed as a fraction of the mean inter-particle separation (commonly $b=0.2$). UNIT scales this with resolution ($b = 0.2\,L_{\rm box}/N_c$ where $\rm L_{ box}$ is the size of the box and $N_c$ the number of cells or particles along one dimension of the simulation). MDPL2 uses a fixed $b=0.2$, providing a standard, resolution-independent definition. The main consequence is that halo masses (and, in some cases, halo sizes) can differ slightly between the two simulations. This effect is most significant for low-mass halos.
    \item \textit{Minimum particles per halo:} UNIT can have halos with as few as 20 particles, including very low-mass systems at the cost of increased noise and less reliable halo properties. In contrast, MDPL2 requires at least 250 particles per halo, so that all identified halos are well resolved and their internal properties are statistically robust, though excluding the low-mass population. 
\item \textit{Force resolution:} Gravitational forces are softened below a characteristic scale to prevent numerical instabilities. UNIT adopts a fixed softening length of $6\,h^{-1}{\rm kpc}$ for all redshifts to preserves a uniform resolution across the simulation. MDPL2 uses a redshift-dependent softening, $13\,h^{-1}{\rm kpc}$ at high $z$ and $5\,h^{-1}{\rm kpc}$ at low $z$ so that halos are reliably detected at both early and late times. Again, the choice of softening primarily affects the smallest halos and their internal structure, particularly at low masses which will be excluded from our analysis.
\end{enumerate}
\end{list}

Overall, the UNIT and MDPL2 simulations reflect different design priorities, particularly in their N-body and \texttt{ROCKSTAR} configurations. The purpose of UNIT is \emph{completeness}, capturing small halos down to dwarf scales, while the goal of MDPL2 is \emph{robustness}, focusing on well-resolved halos. To make the two simulations compatible for our mock catalog construction, we apply a minimum halo mass cut at the \textsc{Agora} limit ${\rm M}_h > 10^{11}~h^{-1}$~M$_{\odot}$ which can equalize their effective resolution. This minimum halo mass cut results in only a $1.32\%$ loss in the \emph{Roman} galaxy sample coming from UNIT. In Appendix \ref{fig:voidsens}, we demonstrate how particle loss affects 3D and 2D voids and show that this cut has a minimal impact on our results, demonstrating that MDPL2 and consequently \textsc{Agora} provides a reliable choice for our mocks.

\section{Sensitivity of void finder to sub/over-sampling of particles}

\begin{figure}[h]
    \centering
    \includegraphics[width=8cm]{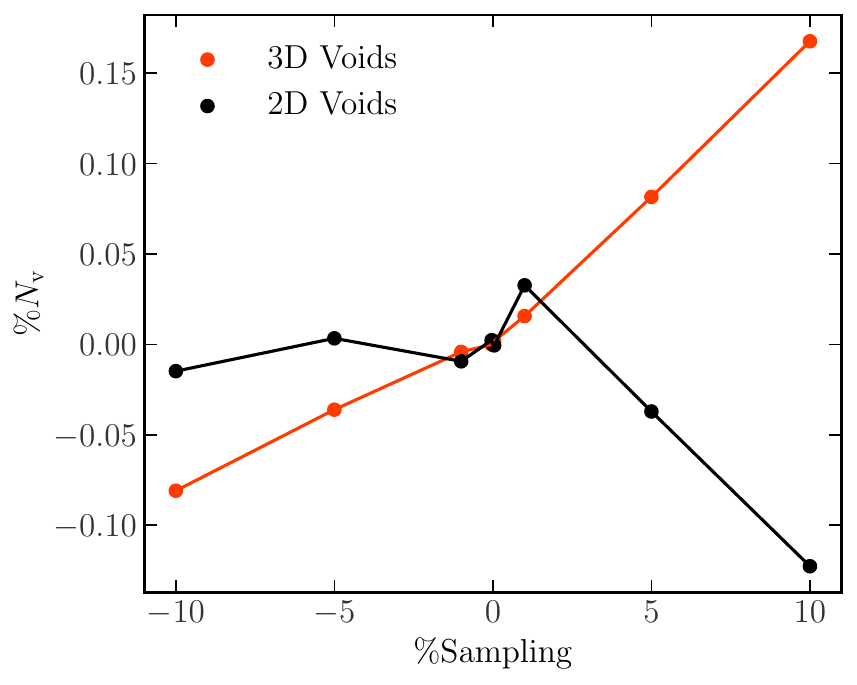} 
    \caption{Sensitivity of 3D and 2D void finders to under- and over-sampling the tracer galaxy field.}
    \label{fig:voidsens}
\end{figure}

Since we have applied a minimum halo mass cut in the reference catalog to equalize the resolution of the simulations, we test the sensitivity of the two used void finders to small changes in the galaxy population. The left side of Fig.~\ref{fig:voidsens} (to the left of the zero on the $x$-axis) represents subsampling of the physical tracer particles in the mock catalog, while the right side illustrates the addition of fake galaxies to the mock catalog.\\

The observed trends depend not only on subsampling or over-sampling but also on the specific void finder used. In the case of subsampling, the 3D void finder detects fewer voids than the nominal case, with the voids being typically larger due to the increase in the volume of voids devoid of matter. In contrast, for the over-sampling case, the random distribution of galaxies causes the fragmentation of some nominal voids into smaller ones.\\

The 2D void finder shows less sensitivity to these changes. When real galaxies are removed, the number of voids remains relatively unchanged. However, when random galaxies are added, the behavior differs from that of the 3D void finder. Due to the 2D void finder’s projected nature, the overall mean density across the map increases, which leads to the identification of fewer voids (only the biggest and deeper ones).

In our case, a decrease of 1.34$\%$ of the tracer galaxies will induce a change in the number of voids of 0.05$\%$ which we consider negligible. \\

In this test, we specifically vary the number density of galaxies, which, as expected, results in changes in the number of voids. What is, however, particularly interesting---and something that was not thoroughly addressed in prior work, except for \cite{tinker2006}---is that even when the number density of galaxies and the power spectra are identical, this does not necessarily lead to the same void statistics as seen in Figure \ref{fig:one_and_two_point_Agora_vel_corrected}. This highlights the need to examine these statistics when generating mock catalogs for void-based analyses or more precise and consistent galaxy analysis.

\section{Different calibration of $M_{*}$ and SFR for \emph{Roman} reference catalog and the \textsc{Agora} simulation}

\begin{figure}[t]
    \centering
    \includegraphics[width=9cm]{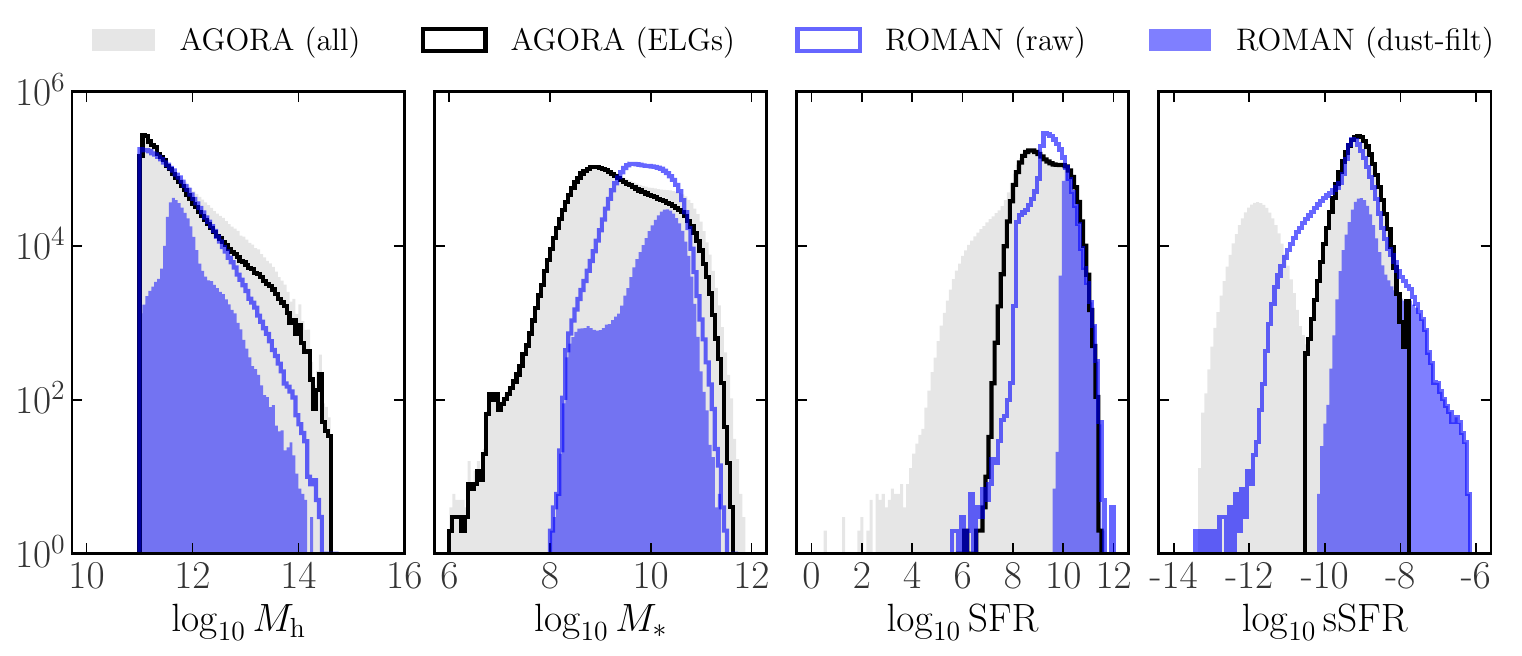} 
    \caption{Distribution of halo mass, stellar mass, star formation rate, and specific star formation rate in \textsc{Agora} (derived from \textsc{UniverseMachine}) and the reference catalog (derived from \textsc{Galacticus}).}
    \label{fig:UM_vs_galacticus_prop}
\end{figure}

In Figure \ref{fig:UM_vs_galacticus_prop}, we compare, for a redshift slice around z = 1.1, the host-halo mass and baryonic properties of \textsc{UniverseMachine} (for \textsc{Agora}) and \textsc{Galacticus} (for the \textsc{Roman} reference catalog).
We present the distributions for the full \textsc{Agora} halo population alongside the flagged ELGs-hosts and compare these to the raw and dust-filtered versions of the \emph{Roman} reference mock catalog. While \emph{Roman} exhibit a slight systematic shift toward higher M$_*$ and sSFR, our analysis demonstrates that these offsets do not significantly impact the \emph{analog matching}'s ability to recover the expected clustering of the \emph{Roman} ELG sample.

\section{Void statistics for 2D void finder with $\mathrm{sm_{VF}} = 5~h^{-1}~$Mpc}

\begin{figure}[h]
    \centering
    \includegraphics[width=9.5cm]{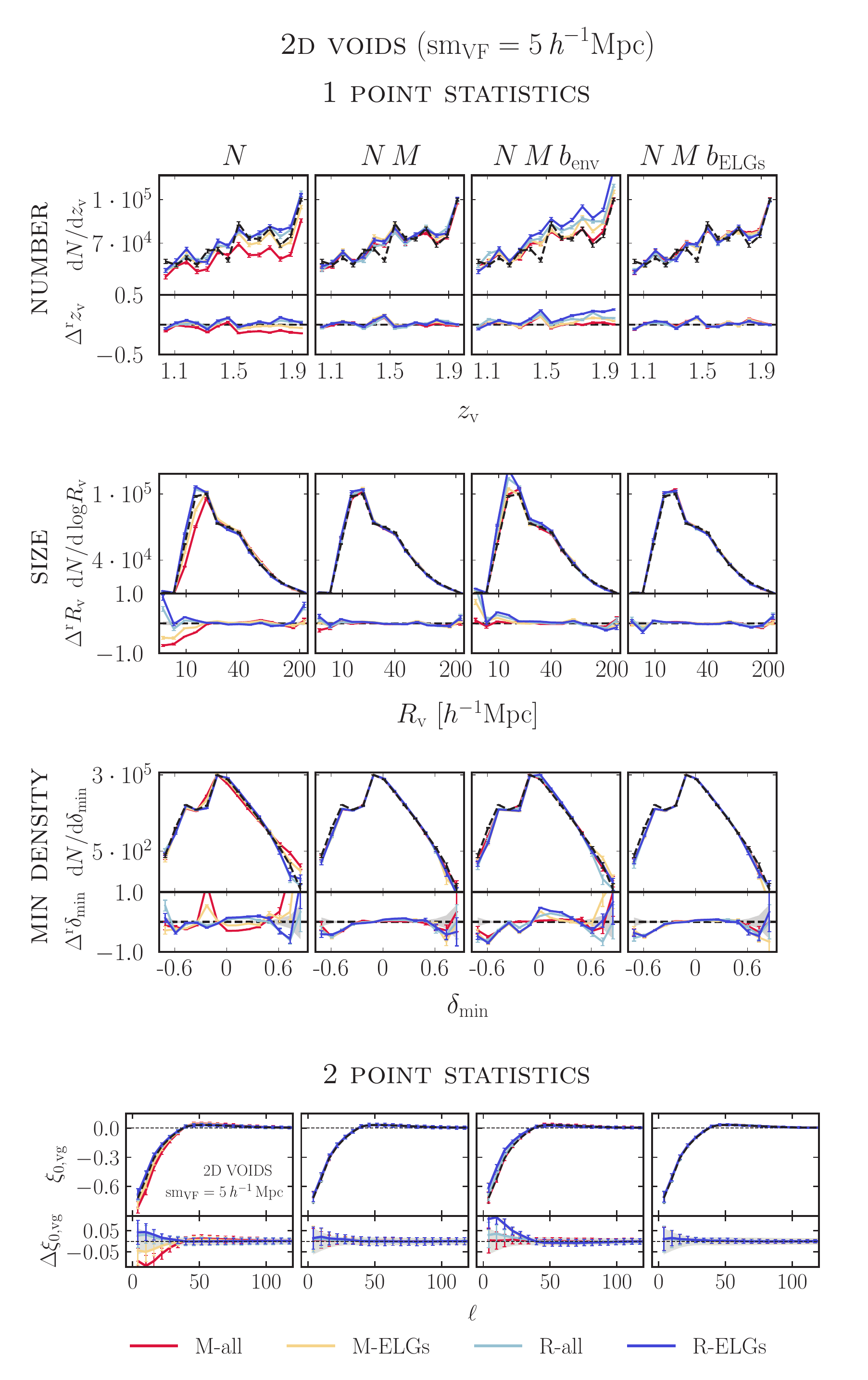} 
    \caption{One- and two-point statistics for 2D voids with $\rm sm_{VF} = 5$~$h^{-1}$~Mpc. Each column represents a different mock family. Different colors denote a different tracer selection. The bottom sub-panels display the residuals, defined as the relative difference with respect to the reference catalog.}
    \label{fig:one_and_two_point_for_2Dvoids_with_sm5}
\end{figure}

Figure \ref{fig:one_and_two_point_for_2Dvoids_with_sm5} shows the one- and two-point statistics for 2D voids identified with a void-finder smoothing of $\mathrm{sm_{VF}} = 5~h^{-1}\mathrm{Mpc}$. Differences between the mock catalogs are more pronounced than in the case of $\mathrm{sm_{VF}} = 10~h^{-1}\mathrm{Mpc}$.

\section{\textsc{Agora} peculiar velocities}

\begin{figure*}[h]
    \centering
    \includegraphics[width=15cm]{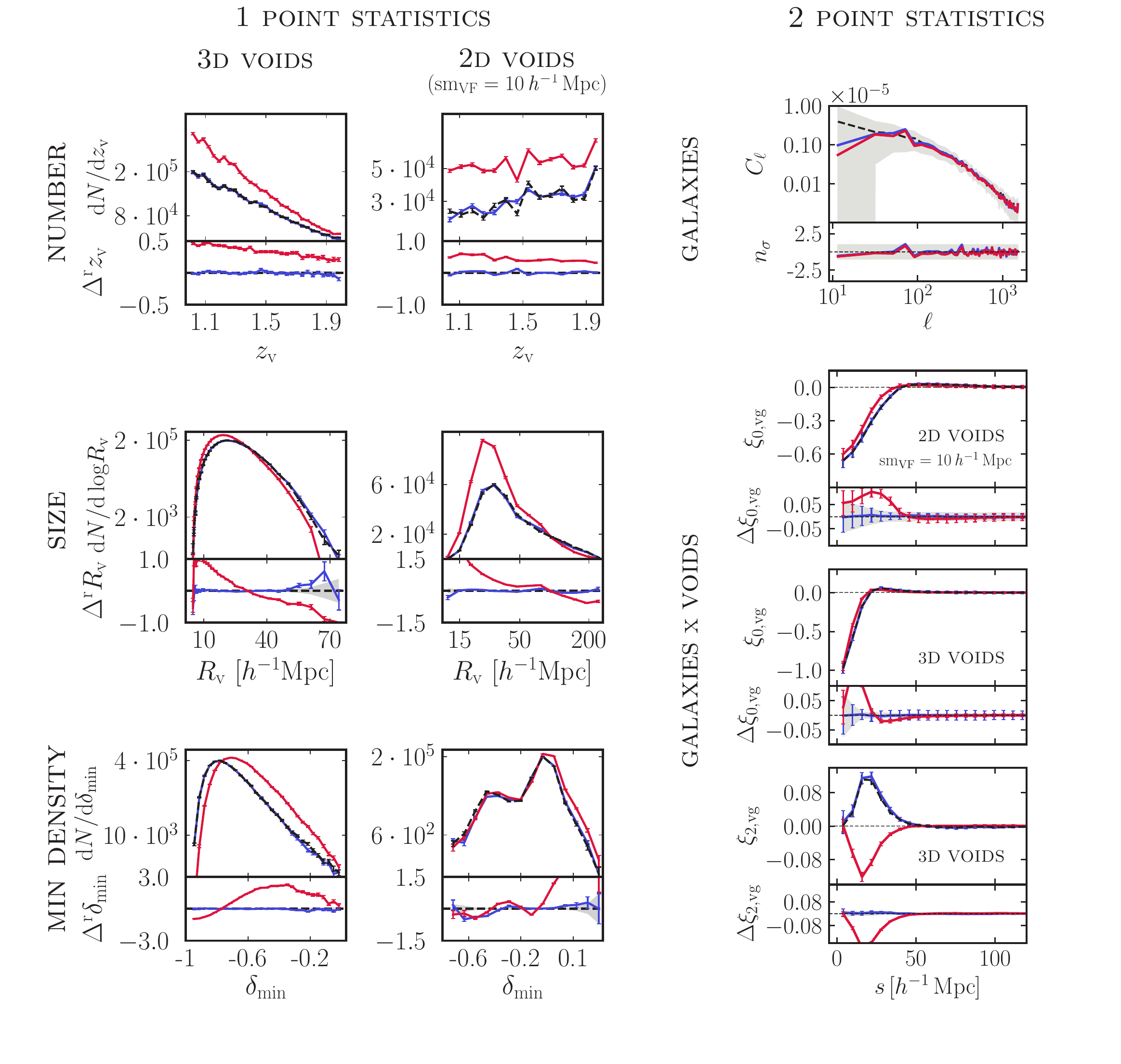} 
    \caption{One- and two-point statistics for 3D and 2D voids ($\rm sm_{VF} = $ 10~$h^{-1}$~Mpc) comparing \textsc{Agora} velocities (original vs. sign-flipped). Residuals relative to \emph{Roman} reference catalog.}
    \label{fig:one_and_two_point_Agora_vel_corrected}
\end{figure*}

We found that the peculiar velocities in the \textsc{Agora} simulation had an unexpected sign inversion. After performing several tests, we confirmed this issue and corrected it by flipping the sign of the velocities in our analysis. Part of those tests consisted on reproducing our findings using the halo lightcone files from \citet{JB2023}, which were independently generated to replicate the \textsc{Agora} lightcone---including the baryonic physics from the \textsc{UniverseMachine}, as we also do---but specifically for Line-Intensity Mapping (LIM) studies. The resulting signals and conclusions using those data sets were consistent with the analyses presented in this work.

\subsubsection{Tests}

\paragraph{Void Size Function and Quadrupole} In the \textsc{Agora} simulation, voids in redshift space appear smaller and more numerous than in real space, in contrast to expectations from the MDPL2 simulation box and other established studies. Previous work shows that redshift-space distortions and the Alcock–Paczyński effect systematically increase void sizes along the line of sight \citep{ryden1996_vsf,lavaux&wandelt2012_stacking_vgcf_aprsd,hamaus2015_vgcf_aprsd, nadathur2019zeldovich,correa2021redshift,correa2022redshift}. These distortions arise from universal physical mechanisms such as peculiar velocities and the cosmological metric, though their magnitude can depend on the simulation’s cosmology ($\Omega_m$, $h$), redshift, and tracer properties (bias, density). However, an inversion where voids shrink at the considered scales and become more numerous is highly anomalous. Figure \ref{fig:one_and_two_point_Agora_vel_corrected} illustrates this behavior: the void size function, monopole, and quadrupole indicate that the original \textsc{Agora} velocities produce voids with apparent matter inflows, which is physically inconsistent.

\paragraph{Angular Redshift Fluctuations} Using angular redshift fluctuations \citet{carlos2019arf}, which are highly sensitive to radial peculiar velocities, we found that theoretical and measured angular power spectra only matched when the sign of $v_{los}$ was inverted.

\paragraph{Independent Cross-Check (\textsc{Skyline})} Using the Skyline code \citet{JB2023}, an independent galaxy catalog generator compatible with \textsc{Agora}'s CMB observables, we confirmed that flipping the line-of-sight velocities was necessary to correctly reproduce the kinematic Sunyaev-Zel’dovich (kSZ) signal of \cite{omori2022}.

\end{document}